 \documentstyle[11pt]{article}
\input epsf
\def\numberbysection{\@addtoreset{equation}{section}
\renewcommand{\theequation}{\thesection.\arabic{equation}}}
\def\theequation{\thesection.\arabic{equation}}

\def\ll{\label}
\def\re{\ref}
\def\c{\cite}

\def\r1{(\ref{$1})}
\def\ot{\otimes}

\def\th{\theta}
\def\ba{\begin{array}{c}}

\def\ea{\end{array}}
\def\pr{\prod}
\def\ni{\noindent}
\def\si{\sigma}
\def\da{\dagger}
\def\De{\Delta}
\def\de{\delta}
\def\bet{\beta}
\def\ov{\over}
\def\ha{{1\over 2}}

\def\l{\left}
\def\l({\left(}
\def\r){\right)}
\def\r{\right}
\def\rw{\rightarrow}

\def\la{\lambda}
\def\al{\alpha}

\def\be{\begin{equation}}
\def\bc{\begin{center}}
\def\ec{\end{center}}
\def\bit{\begin{itemize}}
\def\eit{\end{itemize}}
\def\ee{\end{equation}}
\def\ed{\end{document}}
\def\bea{\begin{eqnarray}}
\def\eea{\end{eqnarray}}
\def\efr{\end{flushright}}

\begin{document}
\title{
Unifying Approaches in  Integrable Systems: Quantum and Statistical,
Ultralocal and Nonultralocal
}

\author{
Anjan Kundu \footnote {
To be published in {\it Classical and Quantum Integrable Systems: Theory and
Applications} (ISBN 07503 09598), Institute of Physics Publishing, (2003)
http://bookmarkphysics.iop.org/
} \\  
  Saha Institute of Nuclear Physics,  
 Theory Group \\
 1/AF Bidhan Nagar, Calcutta 700 064, India.
 \\ email: anjan@theory.saha.ernet.in} 
\maketitle
\vskip 1 cm

\begin{abstract} 
The aim of this review is to present the list of by now a significant
collection of quantum integrable models, ultralocal as well as nonultralocal,
 in a systematic way stressing on their underlying unifying
 algebraic structures. We restrict  to quantum and statistical models
belonging to trigonometric and rational classes with
$2\times 2$ Lax operators. The ultralocal models  can be  classified
 successfully through their associated quantum algebra and are governed by
 the Yang-Baxter equation, while the nonultralocal models, the theory of
 which is still in the developmental stage, allow systematization through
 the braided Yang-Baxter equation. Along with the known integrable models
some  possible directions  for  investigation in this
field and generation of such new models  
  are suggested.
\end{abstract}
\vskip 0.5cm

\noindent
{\bf Key words}: Integrable Quantum and Statistical Vertex Models;
 Quantum Algebras,
Yang-Baxter Equation, Braided Extention,  Algebraic Bethe Ansatz, Ultralocal
and Nonultralocal Models
\section {Introduction}
By quantum integrable systems we will mean the systems with sufficient 
number of higher conserved quantities including the Hamiltonian of the
model. Such a notion of integrability in the   Liouville sense 
allows description through action-angle variables  with the conserved
quantities, which are now operators, playing the role of  action variables. 
For integrable systems the  
 conserved quantities,  being functionally 
independent should form a commuting set of operators
$[c_n,c_m]=0,  \ n,m=1,2, \ldots, N$, such that their total number
 would  match with the
degree of freedom of the system. For example,
  an one dimensional  lattice  model  of $l$-sites describing  $d$-mode
pseudo-particle should have number of conserved quantities $N=d  l$.
Note that for spin-$\ha$ chains we have $d=1$, while spin-$1$ and electron
models account for $d=2$.
 In this review we will stick
to single-mode: $d=1$ systems  for simplicity and consider mainly 
periodic lattice models with $N < \infty$,
 where the algebraic structures can be seen in their exact form.
At the lattice constant $\Delta \to 0$ the field models will be generated
from their exact lattice versions, whenever possible. 
  Integrable field models with $N \to \infty$
 consequently needs to have infinite number of conservation laws.
 
Integrable systems therefore are
  restrictive systems with a very rich
 symmetry. The beauty of such models is that they allow exact solutions for
the eigenvalue problem simultaneously for all conserved operators including
the Hamiltonian. Moreover such $1$-dimensional quantum systems are related
also to the corresponding
$2$-dimensional classical statistical models with a fluctuating variable.
Therefore parallel to a quantum mechanical model one can in principle
exactly solve also a related 
 vertex-type model on a $2$-dimensional lattice using almost the same
techniques and similar results \c{baxter}.
Celebrated examples of such interrelated integrable quantum and statistical
systems are 
$XYZ$ quantum
spin-$\ha$ chain and the $8$-vertex statistical model, 
 $XXZ$ spin chain and the $6$-vertex model, spin-1 chain and the 19-vertex
model etc.

For describing  an integrable system with such an involved structure,
one naturally    can no longer  start  
 from the Hamiltonian of the model as customary in physics,
since now the Hamiltonian is merely one among many commuting
conserved charges. It  therefore  
needs to adopt certain abstractions
 which are formalized by the quantum inverse scattering method and the
algebraic Bethe ansatz (see \c{aba,korbook}).  Though we would use the same
language, we take here a slightly different view point since we intend to
describe integrable systems belonging to both { ultralocal} and {
nonultralocal} classes. For effective description of integrable systems it
is convenient to define a generating function called {\it transfer matrix}
$\tau(\lambda)$, depending on some extra  parameter $\la$ known as 
the  {\it spectral parameter},
 such that one can recover the infinite number of conserved quantities as
the expansion coefficients of $\tau(\lambda)$ or any function of it 
like $\ln \tau(\lambda) = \sum_jc_j\lambda^j$.
The crucial {\it integrability condition}
 may then be defined in a compact form as
\be [\tau(\lambda),\tau(\mu)]=0,\ll{int}\ee
 from which  the commutativity of $c_j$'s follows
immediately by comparing the coefficients of different  powers of $\la , \mu$.

However for solving the eigenvalue problem as well as for identifying the
structure of the model  we require a more general matrix formulation, from
where the integrability condition may be derived. At the same time we
need to transit  from the  global to the local description defined at 
each lattice point, where some individual properties of a model are 
well expressed.
 At this local level, as we  see now, the difference between the ultralocal
and the
 nonultralocal   models become prominent.   
An integrable system allowing the needed  abstraction may be represented by
  an unusual type of matrix called {\it the Lax operator}
$L_{aj}(\la)$ defined at each site $j$ in a 1-dimensional discretized lattice.
The index $a$ defines the matrix or the auxiliary space, while  $j$
designates
the quantum space.
The matrix elements of the Lax operator, unlike in usual matrices
 are operators acting on
some Hilbert space. 
The models with the  Lax operators commuting 
at different lattice sites: 
\be
 L_{aj}(\la)
L_{bk}(\mu)=  L_{bk}(\mu)  L_{aj}(\la), \  a \neq b, \ j \neq k 
\ll{ul}\ee
are known as the  {\it ultralocal }  models, while the integrable models 
for which the above  ultralocality condition does not hold
are classified as the   {\it nonultralocal }  models.
Note that in expressions like (\re{ul}) different auxiliary spaces
   mean different  tensor  products like $L_{1j}(\la)= L_{j}(\la)  \otimes I
$ and $L_{2j}(\mu)= I \otimes 
L_{j}(\mu)  
$. 
 The ultralocal property 
(\ref {ul}) generally  reflects    the  
involvement of canonical operators with 
commutation  relations
 like $[u(x),p (y)]=i \delta(x-y)$ or $[\psi(x),\psi^\dagger (y)]=
 \delta(x-y) $ in the Lax operator   giving
trivial commutator at points $x \neq y$. 
 In nonultralocal models on the other hand the 
basic fields may be  of noncanonical type, e.g. 
$[j_1(x),j_1 (y)]= \delta_x^{'}(x-y)$ or 
 derivatives of the
canonical fields may appear  in their Lax operators 
violating the ultralocal condition and bringing 
  additional complicities, which might not  always be resolved.
Due to this reason the theory and
application for the nonultralocal   models are still in the process of
 development and are far from completion. 
 In spite of   many important models  belonging to this class, it is 
  rather 
disappointing to note that,  this category of models 
 has
  not received  the required attention in the literature.

\section {Integrable structures in  ultralocal models}
\setcounter{equation}{0}
We focus first on the ultralocal systems
due to their relative simplicity  and
formulate a unifying scheme for generating such quantum and statistical
integrable models. 
For ensuring the integrability of an ultralocal model 
it is sufficient to impose certain matrix commutation relation
known as  the {\it quantum  Yang-Baxter equation} (QYBE)
  on its representative  Lax operator in the form
\be
R_{ab}(\la-\mu) L_{aj}(\la)  
L_{bj}(\mu)=  L_{bj}(\mu)  L_{aj}(\la) R_{ab}(\la-\mu),
\ll{ybe}\ee
defined at each lattice site  $j=1,2, \ldots, N.$
 The above QYBE expresses actually  the  commutation relations 
among different matrix  elements of the $L$-operator, given
 in a compact matrix form, where the
structure constants are  determined 
by the spectral parameter dependent c-number elements
 of the $R(\la-\mu) $-matrix. The $R $-matrix in turn  
 should satisfy  a similar but simpler YBE
\be
R_{ab}(\la-\mu)R_{ac}(\la-\gamma)R_{bc}(\mu-\gamma)=
R_{bc}(\mu-\gamma)   R_{ac}(\la-\gamma) R_{ab}(\la-\mu).
\ll{rybe}\ee
Since our intention is to establish the integrability which is a global
property, we have to switch from this local picture at each site $j$ to a
global one by defining a matrix, known as the monodromy matrix  
\be
  T_a(\la)=\prod_{j=1}^N L_{aj}(\la), \ \ \ \ T(\la) \equiv
\left( \begin{array}{c} 
  A(\la), \
  B(\la) \\
  C(\la),  \
  D(\la)
   \end{array} \right). \ll{monod}\ee
 Multiplying therefore the QYBE
(\re{ybe}) for $j=1,2, \ldots, N$ and using the ultralocality condition 
(\re{ul}), thanks to which 
 one can treat the objects at different lattice points 
as commuting objects as in the classical case  
and drag $ L_{aj}(\la)$
 through all $ L_{bk}(\mu) $'s for $k
\neq j, b \neq a$ to arrive  
  at the global QYBE 
\be R_{ab}(\lambda - \mu)~ { T}_{a} (\lambda)~{ T}_{b}(\mu )=  
{ T}_{b}(\mu )~ {
~  T_{a} }(\lambda)~~ R_{ab}(\lambda - \mu).\ll{gybe}\ee
Note that the local and the global QYBE
have exactly the same structural form. Invariance of the  algebraic
form also for the tensor product of the algebras, as revealed here,
indicates the occurrence of the coproduct related to a
deep  Hopf algebra structure  underlying all  integrable systems
\c{frt}.
We will see below that for nonultralocal models such a structure is modified
a bit to include additional braiding relations.
 For the periodic ultralocal models defining further the transfer matrix as 
$~\tau(\la)=tr_a T_a(\la)$, taking $trace$ from both sides of the global YBE
(\ref {gybe}) and canceling  the $R$-matrices due to the cyclic rotation of
matrices under the  trace  we reach finally for $~\tau(\la) $ at the 
trace identity (\re{int})  defining 
the quantum integrability of the system.
Therefore we may conclude that the local  QYBE (\re{ybe})
 in association with the 
ultralocality condition (\re{ul}) is the  sufficient condition for quantum
integrability 
of an ultralocal system.  
Consequently we may  define 
such an integrable system        by its
 representative  Lax operator
together with the associated $R$-matrix  satisfying these criteria.
Note  that  we are concerned 
 here only with the systems with  periodic boundary condition.
  For models with
 open boundaries, the  QYBE  should however be modified with the inclusion
of a reflection matrix, which was introduced in detail 
in   \c{open}.

\subsection {List of well known ultralocal models}
 To have a concrete picture before us,  we  furnish  a list of
 the well known
ultralocal models together  with their $L$-operators and $R$-matrices.
We will however restrict here for simplicity only to the quantum models with
$2 \times 2$-matrix Lax operators  associated with  $4 \times 4$
$R$-matrices.
We  show in the next section how these Lax operators can be generated
in a systematic way confirming their integrability. The 
$R^{\al \beta}_{\gamma \delta}$-matrix 
that satisfies the YBE relation (\re{rybe}), with the indices taking values $1,2$
only,  
can be    given in a simple form
  by defining  its nontrivial elements as
  \c{kulskly}
\begin {equation} 
R^{11}_{11} = R^{22}_{22}= a(\la),
\  R^{12}_{12} = R^{21}_{21}= b(\la), \ R^{12}_{21} = R^{21}_{12}= c 
      .    \ll{R-mat}\end {equation}
 These elements may be expressed explicitly  through 
 trigonometric functions in spectral parameters as 
\be a(\la)=\sin(\la+ \al), \ \ b(\la)= \sin  \la , \ \ 
 c ={ \sin \al} \ll{trm}\ee 
or as its $\al \to 0, \la \to 0$ limit, through rational functions as
\be a(\la)=\la+\al, \ \ b(\la)=  \la , \ \ 
 c ={  \al}. \ll{rrm}\ee
Moreover under a twisting transformation \be 
R(\la) \to \tilde R(\la,\th)=F(\th)
R(\la)F(\th), \ \mbox{ with} \  
F_{ab}(\th)=e^{i\th(\sigma^3_a-\sigma^3_b)} \ll{rtwist}\ee
 one gets twisted trigonometric
and rational $R$-matrix solution of (\re{rybe}) , which may be given by 
 (\re{R-mat})
with the difference $  R^{12}_{12} = b(\la)e^{i\th}, R^{21}_{21}=
 b(\la)e^{-i\th}$.
Apart from these  $R$-matrices there can be 
elliptic $R$-matrix solution, for example that related to the $XYZ$ spin chain
and the 8-vertex model \c{baxter}. All models we  consider here  
 however are associated   with the  trigonometric or the 
rational  $R$-matrices  and in the list presented below 
we group them  accordingly, denoting $ H$ for the Hamiltonian and ${\cal
L} \  (L_n) $ for the Lax operator related to field \  (lattice) models. 
\\
\noindent {\bf I. Models associated with trigonometric $R$-matrix} ($q=e^{i
\al}, \xi=e^{i \la} $)
\\ \noindent
i) {\it Field models}\\ 
1. Sine-Gordon  model \c{sg}
 \begin{equation}
u_{tt}- u_{xx} = \frac {m^2}{\al} \sin(\al u),\qquad 
  {\cal L} = \left( \begin{array}{c} ip , \qquad
  m  \sin (\la-\al u) \\
   m  \sin (\la+\al u),  \qquad -ip
    \end{array} \right), \ \ p= {\dot u}.
\ll{sg}\end{equation}
          \noindent
2. Liouville model  \c{llm}
 \be 
u_{tt}- u_{xx} =  e^{i\al u}, \qquad
  {\cal L}  = i\left( \begin{array}{c} p , \quad
   \xi e^{i\al u} \\
  \frac {1}{\xi}e^{i\al u},  \quad -p
    \end{array} \right), \  [u(x),p (y)]=i \delta(x-y).
\ll{lm}\end{equation}
                     \ni
3. A derivative NLS (DNLS) model \c{qdnls} 
\be
i\psi_{t}- \psi_{xx} + 4i \psi^\da\psi\psi_x=0
, \ \   {\cal L}  = i\left( \begin{array}{c}{
  -\frac{1}{4} \xi^2+k_- N} , \
   {\xi \psi^\da} \\
  {\xi\psi},  \
  {\frac{1}{4} \xi^2-k_+N}
    \end{array} \right), \ \ N=\psi^\da \psi, \ [\psi(x),\psi^\dagger (y)]=
 \delta(x-y)
\ll{dnls}\end{equation}
4. Massive Thirring (bosonic) model (MTM) \c{kulskly} 
$$ H=\int dx \ [ -i \hat \psi^\da ( \sigma ^3 \partial_x + \sigma ^2) \hat \psi
+2 \psi^{(1)\da}\psi^{(2)\da}\psi^{(2)}\psi^{(1)}], \ \hat \psi^\da= 
(\psi^{(1)\da},\psi^{(2)\da}), \ 
[\psi^{(a)}(x),\psi^{\dagger (b)} (y)]=\de_{ab}
 \delta(x-y) $$ 
 \be \   {\cal L}  = i\left( \begin{array}{c}{
 f^+(\xi,N^{(a)})}  , \
   {\xi \psi^{(1)\da}}+{{1 \ov \xi }\psi^{(2)\da}} \\
  \xi\psi^{(1)} +{{1 \ov \xi }\psi^{(2)}}, \ 
  {f^-(\xi,N^{(a)})}
    \end{array} \right), f^\pm (\xi,N^{(a)}) = \pm(\frac{1}{4}( 
\frac{1}{\xi^2}-\xi^2)+k_\mp N^{(1)} -k_\pm N^{(2)})
\ll{mtm}\end{equation}
\noindent
 ii) {\it Lattice Models}  \\ 
1. Anisotropic $XXZ$ spin chain \c{xxz}
\bea
{ H} = \sum_n^N\si_n^1 \si_{n+1}^1+\si_n^2 \si_{n+1}^2 +\cos \al
\si_n^3 \si_{n+1}^3,  \ \
{L_n }(\xi) = \sin(\la + \al \si^3 \si_{n}^3)+\sin \al \
 ( \si^+ \si_{n}^-
+\si^- \si_{n}^+ )
\ll{XXZ}\eea
2. Lattice SG model \c{korepinsg}
\begin{equation}
  L_{n}(\la)  =
  \left( \begin{array}{c}  g(u_n)~ e^{ip_n \Delta },
 \qquad  m\Delta  \sin (\la-\al u_n) \\
   m\Delta  \sin (\la+\al u_n),\qquad   e^{-ip_n \Delta }~ g(u_n)
    \end{array} \right), \quad g^2(u_n)= 1 +  m^2  \Delta^2
\cos  { \al (2u_n+{1 }) } 
\ll{L-sg}\end{equation}
\ni
3. Lattice Liouville model  \c{llm}
\begin{equation}
  L_{n}(\xi)  = \left( \begin{array}{c}   e^{ip_n \Delta }~f(u_n)~,
 \qquad  {\Delta}{\xi}e^{i\al u_n} \\
 \frac{\Delta}{\xi}e^{i\al u_n}   ,\qquad f(u_n)~  e^{-ip_n \Delta }
    \end{array} \right), \qquad f^2(u_n)=
     1 + {\Delta^2} e^{i\al(2u_n+1)}.
\ll{Llm}\end{equation}
\ni
4.  Lattice  DNLS  model \c{construct}
\begin{equation}
  L_{n}(\xi)  =
  \left( \begin{array}{c}
  \frac{1}{\xi}q^{-N_n}- {i \xi\Delta}~q^{N_n+1} ,
 \qquad  {\kappa}A^\da_n \\
  {\kappa}A_n    ,\qquad
  \frac{1}{\xi}q^{N_n}+ {i \xi\Delta}~q^{-(N_n+1)} 
    \end{array} \right), \ [A_n,A_m^\dagger]= \delta_{nm}{\cos \al(2N_n+1) \ov
\cos \al}.
\ll{Ldnls}\end{equation}
\ni 5. Lattice MTM \c{lmtm}

Exact lattice version of MTM (\re{mtm}).

Lax operator: $L_n=
L_n^{(1)} \tilde L_n^{(2)}$ (each factor is a  realization of 
(\re{Ldnls})
for  a  bosonic  mode).  
\\
\ni
6.  Discrete-time  or relativistic   quantum  Toda chain  \c{rtoda}
\begin {equation}
H=\sum_i\left(\cosh 2\al p_i
+\al^2 \cosh \al (p_i+p_{i+1})e^{(u_i-u_{i+1})}
\right), \ \ 
L_n(\xi) = \left( \begin{array}{c}
  \frac {1}{\xi}e^{\al p_n}
-\xi  e^{-\al p_n}, \ \ \  \al e^{u_n}
 \\-\al e^{- u_n
}
 ,\qquad  0
          \end{array}   \right).
\ll{rtodal}\end {equation}
\noindent{\bf Ia. Models associated with twisted trigonometric  $R$-matrix
 }\\
\ni
6.a) Quantum Suris discrete-time  Toda chain \c{suris,rtoda}
 \begin {equation}
L_k(\xi) = \left( \begin{array}{c}
  \frac {1}{\xi}e^{2\al p_k}
-\xi , \ \ \  \al e^{u_k}
 \\-\al e^{2\al p_k- u_k
}
 ,\qquad  0
          \end{array}   \right),
\ll{lsuris}\end {equation}
\ni
7. Ablowitz-Ladik model \c{al,kulskly}
\be  ib_{j,t}+(1+\al b^\dag_jb_j)(b_{j+1}+b_{j-1})=0, \ \ L_{k}(\xi)  =
  \left( \begin{array}{c}
  \frac{1}{\xi},
 \qquad b^\dag_k \\
  b_k    ,\qquad
  {\xi}
    \end{array} \right), \ [b_k,b^\dag_l]=\de_{kl} (1- b^\dag_k b_k)
 \ll{lal}\ee
\noindent{\bf II. Models associated with rational $R$-matrix
 }\\
\noindent
i){\it Field models:} \\
 1. Nonlinear Schr\"odinger equation (NLS)
\begin{equation}
i\psi_{t}+ \psi_{xx} +  (\psi^\da\psi)\psi=0,
      \quad
{\cal L}(\la)  = \left( \begin{array}{c} \la ,
\quad  \psi \\
\psi^\da, \quad -\la
    \end{array} \right).
\ll{nls}\end{equation}
\noindent
 ii) {\it Lattice Models:}   \\
1. Isotropic $XXX$ spin chain \c{xxz}
\be
{ H}= \sum_n^N \vec \si_n\cdot \vec \si_{n+1} , 
 \  \ \
L_{an}(\la) =  \la {\bf I} + {\al } P_{an}, \ \  P_{an} ={1 \over 2} 
({\bf I} + \vec \si_a\cdot \vec \si_{n})
\ll{XXX}\ee
\ni
2. Gaudin model  \c{gaudin}

In the simplest case the Hamiltonians 
\be H_k= \sum_{l \neq k}^N {1 \ov \epsilon_k-\epsilon_l} (\vec \sigma_k \cdot
 \vec \sigma_l), \ \ k=
1,2, \ldots, N , \ \ \ 
\ \ \  L_{ak}(\la)= (\la - \epsilon_k) {\bf I} + {\al } P_{ak}.
\ll{gaud}\ee
\ni
3. Lattice NLS  model  \c{korepinsg}
\be L_n(\la) = \left( \begin{array}{c}
\la +s- \De N_n   \qquad    \De^\ha(2 s- \De N_n)^\ha\psi_n^\da
 \\
\De^\ha\psi (2 s- \De N_n)^\ha 
 \qquad \quad \la -s +\De N_n
          \end{array}   \right), N_n=\psi_n^\da \psi_n, \  [\psi_k,\psi^\dag_l]=\de_{kl}.
\ll{lnls}\end {equation}
\ni 4. Simple lattice NLS \c{kunrag}
\be L_n(\la) = \left( \begin{array}{c}
\la +s -N_n\qquad    \psi^\dag_n
 \\ \psi_n
 \qquad \quad -1
          \end{array}   \right).
\ll{snls}\end {equation}
\ni
5. Discrete self trapping dimer model \c{enolskii91}
\be H=-\left[\ha \sum_a^2 (s_a -N^{(a)})^2+(\psi^{\dag(1)}\psi^{(2)}+
\psi^{\dag(2)}\psi^{(1)})\right], [\psi^{(a)},\psi^{\dag(b)}]=\de _{ab},
 a,b=1,2 \ll{dst}\ee
Lax operator $L(\la)=
L^{(1)}(\la)L^{(2)}(\la),$  ( each factor 
as (\re{snls})
for each of two bosonic modes).  
\\ \ni
6. Toda chain    (nonrelativistic) \c{kulskly} 
\be
H=\sum_i\left(\ha p^2_i
+ e^{(u_i-u_{i+1})}
\right),\qquad
L_n(\la) = \left( \begin{array}{c}
  p_n
-\la \qquad    e^{u_n}
 \\- e^{-u_n
}
 \qquad \quad 0
          \end{array}   \right).
\ll{toda}\end {equation}

\section{  Unifying algebraic approach in ultralocal models}  
\setcounter{equation}{0}
Though the QYBE itself represents  an  unifying approach
for all ultralocal models, we     intend to specify here 
a common  algebraic  structure independent of the spectral parameter 
that will 
 not only systematize  the models including those 
listed above,
  but also identify 
 their  common integrable  origin,   
  establishing  naturally the quantum integrability 
for all of them,  simultaneously.
From  the above list of models, 
  one may observe that,  
 different   integrable 
models have  their representative  Lax operators 
 in diverse
 forms with varied  dependence on  the spectral parameter
as well as
 on the basic operators like  
  spin, bosonic  or
the canonical  operators. However 
the $R$-matrices  associated with
all of them 
 are given by  the
 same   form (\re{R-mat}) with known trigonometric (\re {trm}) or 
its limiting rational (\re {rrm}) solutions. 
To explain this  intriguing  observation 
we may look for a common origin for the Lax operators
linked with a general 
underlying algebra free from spectral parameters, though derivable 
 from the QYBE.
 We propose to take the  Lax operator of such   ancestor
model   in the form \c{kunprl99}
\be
L_{trig}^{(anc)}{(\xi)} = \left( \begin{array}{c}
  \xi{c_1^+} e^{i \al S^3}+ \xi^{-1}{c_1^-}  e^{-i \al S^3}\quad \ \ \ 
\epsilon_+  S^-   \\
    \quad  
 \epsilon_- S^+    \quad \ \ \ \xi{c_2^+}e^{-i \al S^3}+ 
\xi^{-1}{c_2^-}e^{i \al S^3}
          \end{array}   \right), \quad
 \xi=e^{i \alpha \la}, \ \epsilon_\pm =2 \sin \al \xi ^{\pm 1}, \ll{nlslq2} \ee
where $\vec S$ and $ c_a^\pm,\  a=1,2$ are some operators, the algebraic 
 properties of 
which are specified below. The 
structure of (\re {nlslq2})  becomes  clearer if we notice
the decomposition 
$ L_{trig}^{(anc)}{(\xi)}= \xi L_++\xi^{-1} L_-$, where $L_\pm$ are 
spectral parameter $\xi$-independent upper and
lower triangular  matrices similar to the construction of \c{frt}.
 Inserting (\re {nlslq2}) in QYBE  together with its associated 
 $R$-matrix (\re {R-mat}) with  trigonometric solution  (\re {trm})
 and matching different powers of the
 $\xi$ we obtain the underlying general algebra
as
\be
 [S^3,S^{\pm}] = \pm S^{\pm} , \ \ \ [ S^ {+}, S^{-} ] =
 \left ( M^+\sin (2 \al S^3) + {M^- } \cos
( 2 \al S^3  ) \right){1 \over \sin \al}, \quad  [M^\pm, \cdot]=0,
\ll{nlslq2a}\ee
with  $ M^\pm=\pm \ha   \sqrt {\pm 1} ( c^+_1c^-_2 \pm
c^-_1c^+_2 ) $ behaving as central elements with arbitrary values of
$c$'s.  As we have mentioned above, the integrable systems are associated
 with an important  Hopf algebra $A$,
 exhibiting the
properties like  
1) {\it coproduct}  $\Delta(x): A \rw A \otimes A$, 2) {\it antipode} or
'inverse' $S:A \rw A  $,
3) {\it  counit} $\epsilon: A \rw k, $   4)  
{\it multiplication}
 $ {\sl M}: A \otimes A \rw A $ and  5) {\it unit} $ \al : k \rw
A $. It can  be shown that all these properties   hold also  for     
(\re{nlslq2a}) defining it 
 as a  
Hopf algebra. Referring the interested readers to the original works
\c{qa} for more mathematical
treatment  of
 the noncocommutative Hopf algebra, we give here only some  
simple and
intuitive  arguments in its
constructions. For example  
 the coproduct $\Delta(x)$, the most important of these 
characteristics, can be derived for algebra (\re{nlslq2a}) 
by  exploiting    a    QYBE   property  that
  the  product of
two Lax operators  $L_{aj}L_{aj+1}$ is again 
a  solution of the  QYBE and may be given in the explicit form
as
\bea
\Delta(  S^+)=c_1^{+ }e^{i \al S^3} \otimes S^+
                 +   S^+ \otimes c_2^{+ }e^{-i \al S^3} &
,&  \Delta(   S^-)= c_2^{- }e^{i \al S^3} \otimes  S^- +
   S^- \otimes c_1^{- }e^{-i \al S^3}\nonumber \\
\Delta( { S^3})= I \otimes { S^3}+S^3 \ot I
&,& 
\Delta(  c_i^{\pm })=c_i^{\pm } \otimes c_i^{\pm }. 
\ll{Detsa}\eea
The multiplication property mentioned above is also in agreement with the 
ultralocality condition, which is  used for transition from local 
to global QYBE
following the multiplication like 
\be 
({A}\otimes  {B})(C \otimes D) = (AC \otimes BD) \ \mbox{with} \
 A=L_{i}(\la),
 B=L_{i}(\mu),
 C=L_{i+1}(\la),
 D=L_{i+1}(\mu). \ll{mult}\ee  
Note that (\ref{nlslq2a}) is a $q$-deformed 
  algebra and a generalization of the 
well  known quantum algebra \c{qa} $U_q(su(2))$.

In fact different  choices
of  the central elements $c^\pm_a$ reduce  this algebra to the 
q-spin, q-boson as well as various other 
q-deformed algebras along with their undeformed limits. Therefore we can
 obtain
easily the coproduct for these algebras, whenever
admissible, 
 from their general form (\re{Detsa}) in a systematic way 
by taking  the corresponding
values of $c'$s. 
\subsection {Generation  of models }
We know that the well known integrable models listed above were 
 discovered  at different points of time,  
mostly in an isolated way  and
   generally by
quantization of  the existing classical models.
However, as we will
see,  they can actually 
 be generated in a systematic way 
through various  realizations of the same Lax operator (\re
{nlslq2})  giving  a unifying picture of integrable ultralocal models.
For this we find 
first a representation of (\ref{nlslq2a}) like
\be
 S^3=u, \ \ \   S^+=  e^{-i p}g(u),\ \ \ 
 S^-=  g(u)e^{i p}, 
\ll{ilsg}\ee
 in physical variables
with $  [u,p]=i,$
where
the operator function
\be g (u)= \left ( {\kappa  }+\sin \al (s-u) (M^+ \sin \al (u+s+1)
+{M^- } \cos \al (u+s+1 ) ) \right )^{\ha}  { 1 \ov \sin \al }, \ll{g}\ee
containing free parameters $\kappa$ and $s$. We demonstrate now  that
 the  Lax operator (\ref{nlslq2}), which 
represents a {\it generalized lattice  SG } like model for 
(\re{ilsg}) 
   may  serve as  
an ancestor model (with possible 
realizations also in other physical variables like 
  bosonic 
  $  \psi, \psi^\dag
$ or  spin  $s^\pm,s^3$ operators) 
 for generating all   integrable ultralocal quantum as well as 
 statistical systems.
As an added advantage,  the  Lax operators of these 
 models are derived 
automatically from (\ref{nlslq2}), while the
$R$-matrix is  simply inherited.  The underlying algebras of the models
are also given by
the corresponding representations of   
the ancestor algebra  (\ref{nlslq2a}), which being a direct consquence  of
 the QYBE
ensures  the quantum integrability of all its descendant models, that we
 construct here. It should be stressed that due to the symmetry of
the solution (\re{R-mat}): $ [R(\la-\mu), \sigma^a \otimes \sigma^a]=0,
\ a=1,2,3$ the
Lax operator   (\ref{nlslq2}) as a solution of QYBE 
may be right or left multiplied  by any 
$\sigma^a$. We shall use this freedom in our following  constructions, 
whenever needed.

Note that we may  generate also the 
quantum field models  by taking properly the 
continuum limit   of  their lattice variants with
the  lattice spacing $\Delta \rw 0 $. Though in general such transitions to
the  field limit might be tricky and problematic we suppose  their 
validity  
assuming  the lattice  operators 
 to go smoothly to
 the field operators $p_j \rw p(x), \psi_j \rw \psi(x)$, 
 with the corresponding commutators:   $ [\psi_j,\psi_k^\dag]= {1 \ov \Delta}\delta_{jk}
 \ \rw \
 [ \psi(x),\psi(y)]=\delta(x-y)$ etc. 
The lattice Lax operator therefore should reduce to its field counterpart
 ${\cal L}(x, \la)$
 as $L_j(\la) \rw I+ i\De {\cal L}(x, \la) +O(\De^2).$
 The associated $R$-matrix however remains the same, since it does not contain
lattice constant $\De$. Thus {\it integrable field models} like sine-Gordon,
 Liouville, NLS or the derivative NLS models can be recovered
 from their
exact lattice versions and having the same quantum $R$-matrix,
 though all discrete models may not
always have such a direct field limit.
\subsubsection{Models belonging to trigonometric class}
1.) Choosing trivially all central elements  as $c^\pm_a =1,\ a=1,2$, which 
gives
  $ M^-=0, M^+=1, $   
 (\ref{nlslq2a}) reduces clearly  
  to the well known quantum algebra
 $U_q(su(2))$   \c{qa} given by
\be
 [S^3,S^{\pm}] = \pm S^{\pm} , \ \ [ S^ {+}, S^{-} ] =  [{2} S^3]_q.
\ll{slq2}\ee
with the known form of its  coproduct recovered  easily from 
(\re{Detsa}).
 The simplest representation ${\vec S} = 
\ha  {\vec \si}$  for this case derives from  (\ref{nlslq2})
 the integrable $XXZ$ {\it spin chain} (\re{XXZ}). On the other hand,
representation   (\ref{ilsg}) with the corresponding reduction of  (\re{g})
as
$ g (u)={1 \ov 2 \sin \al}
  \left [ 1+ \cos \al (2 u+1)
 \right ]^\ha $ with suitable choice of parameters $s,\kappa$ 
recovers the    Lax operator of
   {\it lattice sine-Gordon} model (\re{L-sg})  
   directly  from (\ref{nlslq2}) and at its field limit the 
field Lax operator (\re{sg}). Note that the spectral dependence in 
$\epsilon_\pm$ appearing in (\ref{nlslq2}) can be easily removed through a
simple gauge transformation \c{frt} and therefore we 
 ignore them in our construction and  use the freedom of 
translational symmetry of the spectral parameters $\la \to \la +const.$,  
whenever needed. 

2.)
 An  unusual
 { exponentially} deformed  
 algebra can  be
generated from (\ref{nlslq2a}) 
  by
fixing the elements as 
$ \ c^+_1=c^-_2=1, \ \ \ c^-_1=c^+_2=0,$ which 
gives  $ M^\pm=\pm \ha \sqrt {\pm 1}$ and

  \be
 [S^3,S^{\pm}] = \pm S^{\pm} , \ 
[ S^+, S^-]= {e^{2i\al S^3} \over 2 i \sin \al } \ll{expa} \ee
and 
reduces  (\re{g}) to 
$  g(u)= {(1+e^{i \al(2 u+1)})^\ha \ov \sqrt {2} \sin \al} $. 
  This  algebra and the
corresponding realization 
 yields clearly from (\ref{nlslq2})  the  Lax operator of the {\it  lattice 
 Liouville } model (\re{Llm}) and at its field limit that of the Liouville
field model (\re{lm}).

It is interesting to observe here that 
though the underlying algebraic structure and hence its realization are
fixed by the choice of  $ M^\pm$, the Lax operator (\ref{nlslq2}) depends
explicitly on the set of $c'$s and therefore may take different forms 
for the same model. For example in the 
  present case   
$ \ c^-_1 \not =0$ would  give again the   
  same  value for $ M^\pm$
but a different 
  {\it  Liouville Lax
operator}  
\c{fadliu} more convenient for the  Bethe ansatz solution.

This opens up  therefore  interesting
possibilities for obtaining  systematically
different useful Lax operators  for the same
 integrable model, as well as for  
constructing  new nonultralocal models \c{kun02}.

3.) 
Recall that the  well known 
$q$-bosonic algebra may be  given by \c{qoscl} 
$ [A,N] = A, \ \  [A^\dag,N] =- A^ \dag,\ \ \  
AA^ \dag- q^{-2} A^ \dag A= q^{2 N} 
$
or in its conjugate  form with $q \to q^{-1}$. Combining these two forms 
we can easily write the commutator of such q-bosons as 
\be
 [A,N] = A, \ \  [A^\dag,N] =- A^ \dag,\ [ A, A^ \dag ] = {\cos (\al (2N+1))
\ov \cos \al} .
\ll{qola}\ee
It is interesting to find that 
 for the  choice of the central elements  
$ \ \ c^+_1=c^+_2=1, \ c^-_1= -{iq }  , \ c^-_2=  {i \over  q} 
$
compatible with   $ M^+=2 {\sin \al } 
, \ M^-= 2i {\cos \al }  $ we may get  a
 realization 
\be S^+= -\kappa A, \ S^-= \kappa A^\dag, \ S^3= -N, \ \
 \kappa= -i (\cot \al)^\ha
,\ll{sqosl}\ee 
  with  (\ref{nlslq2a})  reducing  directly  to the  relation
(\re{qola}), which  
gives   thus  
 a new  integrable
 {\it $q$-boson model}. It is important  to note now that
either   using  (\re{ilsg}) which    simplifies   (\re{g}) to $g^2(u)=  
{[-2u]_q}$ or directly taking the mapping
of the $q$-bosons   to standard bosons:
  $ A =
     ~\psi (\frac{[2N]_q}{2N \cos \al})^{\ha}, \ N=
     \psi^\da\psi \ , $ 
we may  convert  (\ref{nlslq2}) with (\re{sqosl}) to  
an exact lattice version  
 of the quantum {\it derivative nonlinear Schr\"odinger} (QDNLS) equation
(\re{Ldnls}) and consequently  to the  QDNLS field model 
(\re{dnls}).
The QDNLS is 
 related also to the interacting bose gas with derivative
$\delta$-function potential \c{shirman}.

4.) Since the matrix product of Lax operators with each factors
 representing different Lax operator realization for the
same model   
should give again a QYBE solution, we can construct multi-mode integrable 
extensions by taking the product of single-mode Lax operators. Using this
trick, i.e. by  combining two QDNLS models constructed above as
 $L(c^\pm_1, c^\pm_2, \psi^{(1)}) 
L(c^\mp_2, c^\mp_1, \psi^{(2)}) =L(\la)$,   
 we can
 create  further  an integrable exact
 {\it lattice version of the massive Thirring}  model \c{lmtm}.  
At the continuum limit it 
goes to the
 {\it bosonic massive Thirring}
 model introduced in \c{kulskly}, the field Lax operator (\re{mtm})
 of which can be
given simply by the superposition 
${\cal L}={\cal L}^{(1)}(\xi, k_\pm, \psi^{(1)}) +\sigma^3{\cal L}^{(2)}
({ 1\ov
\xi}, k_\mp, \psi^{(2)})\sigma^3 $,  where $1\pm ik_\pm \sin \al=e^{\pm i{\al
\ov 2}} $ and the constituing operators ${\cal L}^{(a)}$
   is given clearly by the DNLS Lax operator
(\re{dnls}) for each of its two bosonic modes.

5.) Since the  general algebra permits  trivial
eigenvalues for central elements, one may choose  both
   $M^\pm=0,$ which  might correspond to  
  different sets of choices  like
$ i) \ \ c_a^+=1 \ , a=1,2, \ \mbox{or} \ ii) \ \ c_a^-=1 \ , a=1,2,
 \ \mbox{or} \ iii)\ \  c_1^\mp=\pm 1  , \  \ \mbox{or} \ \
  iv)\ \  c_1^+=1, \
$
with   the rest of  $c'$s being zero.   
It is easy to see that all of these sets  
 lead to the same  underlying  algebra
\be
[ S^+, S^-]= 0, \ [ S^3, S^\pm]= \pm S^\pm ,\ll{nula} \ee
though   generating   different   
  Lax operators  from (\ref{nlslq2}).

As here   (\re{g}) gives 
simply  $g(u)=$const., 
 interchanging canonically  $u \rw -ip, p\rw -i u,$
  from (\ref{ilsg}) one gets 
\be
 S^3 =-ip, \ S^\pm=  \al e^{\mp u } \ \ 
, \ll{dtl1} \ee
which evidently generates   from the same 
  general Lax operator  (\ref{nlslq2})  
     the  { discrete-time or {\it  relativistic
quantum Toda chain} (\re {rtodal}).
  Note that, 
 iii) and iv)   give two different  Lax operators  found in
\c{rtoda} and \c{hikami} for  the  { relativistic 
 Toda chain}.  Case i) and ii) 
on the other hand could be used for constructing nonultralocal
quantum models, namely   
   { light-cone SG} and the mKdV model \c{kun02}.

 {\bf Models in twisted trigonometric class}: 
Under twisting when the $R$-matrix changes as (\re{rtwist})  
the associated Lax
operator  is also transformed similarly as    
$ L_n(\la) \to \tilde L_n(\la,\th)=F_n(\th)
L_n(\la)F_n(\th),$ \ { with} \  
$F_{an}(\th)=e^{i\th(\sigma^3_a-S^3_n)}$. As a result the ancestor model
(\re{nlslq2})  associated with the trigonometric twisted $R$-matrix (\re{rtwist})
 gets deformed with its operator elements changing as
\be
c^\pm_a \rightarrow c^\pm_a e^{ -i\th S^3_k}, \ \  \  
S^\pm_k \rightarrow 
\tilde S^\pm_k= e^{-i\ha \th S^3_k}
~ S^\pm_k ~ e^{-i \ha \th S^3_k},
 \ll{etwist}\ee
and as a consequence the diagonal elements of the twisted Lax operator take
the form $e^{i(\theta\pm \al)S^3_k}$, with obvious preferance for the
choice $\th =\pm \al$.

6.)  We may  generate the {\it quantum analog of Suris
 discrete-time Toda chain}
 belonging to the twisted class by starting from the ancestor model
  with the change (\re{etwist}), but by fixing the parameter $\th=-\al$ (an
equivalent model is obtained by the choice $\th=\al$). Using the same 
realization (\re{dtl1}) we arrive now at the explicit form (\re{lsuris}).

7.)  However  if we start from
the same twisted ancestor model 
with the same value $\th=-\al$ of the twisting parameter, 
but take the central elements as   
$c^+_1=c^-_2=0$ with $c^-_1=c^+_2=1$ giving 
$ M^\pm= \ha   \sqrt {\pm 1}$
 (compare  with the Liouville case!), 
 all  noncommuting operators   clearly vanish
 from
the diagonal elements of 
 the resulting Lax operator. Moreover 
renaming  the deformed operators as
 $ b_k=2 \sin \al \tilde S^+_k$, 
we get their modified  algebra as  a type of $q$-boson :
 $~~ [b_k,b^\dag_l]=\de_{kl} (1- b^\dag_k b_k) ~~$ and thus generate 
finally the 
exact form of the  {\it Ablowitz-Ladik model} ({\re{lal}).

The domain of the models considered  can therefore be considerably 
extended if we use twisting and some other allowed transformations
\c{wadati} that preserves integrability.

\subsubsection{Models belonging to the rational class}
 One of the crucial parameters inbuilt in both $R$-matrix and  above Lax
operators  is
  the deformation parameter  $q=e^{i \al}$, the  physical meaning of which is
   as 
 anisotropic  or relativistic parameter.
We consider now 
  the undeformed 
limit  $q \rw 1$ or    $\al \rw 0$ 
 related to    isotropic or  nonrelativistic models
belonging to the  rational class, which reduces
 various $\al $-dependent objects  as
 $S^\pm \rw  is^\pm,\{ c^\pm_a\} \rw \{ c_a^i\}, \
M^+ \rw -m^+, M^- \rw -\al
m^-,  \ \xi \rw 1+ i \al \la $. This  transforms  
  (\ref{nlslq2a}) to a $q$-independent algebra
with 
\be  [ s^+ , s^- ]
=  2m^+ s^3 +m^-,\ \ \ \ 
  ~ [s^3, s^\pm]  = \pm s^\pm, 
  \ll{k-alg} \ee
were $m^+=c_1^0c_2^0,\ \  m^-= c_1^1c_2^0+c_1^0c_2^1$ and  
$  c_a^i, i=0,1$ are 
  central to  (\re{k-alg}).
 Note that (\re{k-alg}) is a generalization of  spin 
as well as the bosonic algebra and its coproduct can be obtained as a limit
of (\re{Detsa}).
Consequently, the general Lax operator   (\ref{nlslq2})
   is converted into  
 \be
L^{(anc)}_{rat}{(\la)} = \left( \begin{array}{c}
 {c_1^0} (\la + {s^3})+ {c_1^1} \ \  \ \quad 
  s^-   \\
    \quad  
s^+    \quad \ \ \ 
c_2^0 (\la - {s^3})- {c_2^1}
          \end{array}   \right), \ll{LK} \ee
  and 
   the quantum
 $R$-matrix (\re{R-mat}) is reduced   to its rational form (\re{rrm}).

We would see  that the ultralocal integrable systems belonging to the 
rational class can be generated in a similar way  now
 from the   Lax operator (\ref{LK})  with  algebra (\ref{k-alg}), 
 all sharing
the same rational $R$-matrix (\re{rrm}).
It is not difficult  to check by a
 variable change $(u,p) \to (\psi, \psi^\dag)$
  that
 at the limit $\al \rw 0$  (\ref{ilsg})
 reduces
 to a generalized Holstein-Primakov transformation (HPT)
 \be
 s^3=s-N, \ \ \ \    s^+= g_0(N) \psi, \ \ \  s^-= \psi^\dag g_0(N)
, \ \ \ \ g_0^2(N)=m^-+m^+ (2s -N), \ \ N=\psi^\dag \psi,
\ll{ilnls} \ee
which is also
  an exact realization of  (\ref{k-alg}). Therefore
  Lax operator (\ref{LK}) with such a realization 
    may be considered as
 a {\it generalized  lattice NLS}, which  would
 serve  as a generating model for  all   
  quantum integrable  models belonging to the rational class.

1.)  The   choice
 $  m^+  =  1,m^-  =  0, $
clearly  reduces (\ref{k-alg})    to  
   $su(2)$ algebra
$ \  [ s^+ , s^- ]
=  2 s^3 ,\  \ 
  ~ [s^3, s^\pm]  = \pm s^\pm.
\ $
A  compatible choice  $ c^0_a=1, c^1_a=0$ 
   yields   from (\re{LK}) for the
  spin-$\ha$
representation the Lax operator of
 the  $XXX$ {\it  spin chain} (\re{XXX}).

Taking spin-$\ha$ and spin-$1$ realizations alternatively along the lattice
we can construct  now 
the integrable alternate spin model discovered in \c{alternates}.

Note that  a slightly different choice 
  $ c^0_1=- c^0_2=1, c^1_a=0$ giving $  m^+  =  -1, m^-  =  0, $  
generates  on the other hand the corresponding model with  $su(1,1)$ algebra.

The bosonic realization (\ref{ilnls}) in present cases with 
$  m^+  =  \pm 1, m^-  =  0, $  is simplified 
 to the standard  HPT
with  $ \ \ g_0^2(N)= \pm(2s -\psi^\dag \psi) $, which reproduces
  from (\re{LK})    
the exact {\it lattice   NLS} model  (\re{lnls}) and 
at the continuum limit the more familiar 
   {\it NLS field model} (\re{nls}), with  $+ (-)$ sign in the HPT
 corresponding 
to the attractive (repulsive) interaction. 

2.) A complementary  choice
 $  m^+  =  0,m^-  =  1, $ on the other hand  converts (\re{ilnls}) to
$s^ {+}=\psi, s^ {-}=\psi^\dag, s^ {3} = s -N $
due to  $g_0(N)=1$ and reduce
  (\ref{k-alg}) directly to the standard bosonic relations 
$ \
 [\psi ,N] = \psi,\ \  [\psi^\dag,N] = -\psi^ \dag,\  \ 
 [ \psi, \psi^ \dag ] = 1 $.
 Remarkably,  (\ref{LK}) with this realization 
  generates yet another {\it simple   
lattice NLS} model with Lax operator 
(\re{snls}). 

3.)  Combining two such bosonic Lax operators 
 (\re{snls}), constructed above: $L^{(1)}(\la)L^{(2)}(\la)=L(\la) $   
and considering them to be inserted  at a single site we can construct the Lax
operator of  an integrable  model
involving two-bosonic 
modes, which yields the {quantum discrete self trapping model}
(\re{dst}).  

4.)
Note that  the trivial choice $m^\pm  = 0$  gives again 
  algebra (\ref{nula}) and  hence the 
  realization  (\ref{dtl1}). This  however  yields  
  from (\ref{LK})  the Lax operator  of 
      the  {\it nonrelativistic
 Toda chain}  (\re{toda}) 
  associated  with    the 
 rational  $R$-matrix. 
It is interesting  to note that  in   \c{tarasov}
  the  Lax operators like (\ref{nlslq2}) and
({\ref{LK}) appeared  
 in their bosonic realization and  were 
  shown  to be  the most general  possible form 
within their respective class.

 Therefore these quantum Lax operators
 are in the core of the ultralocal  integrable
models,  
 both discrete and continuum, which
   can be constructed from them in a unified way. 
    Models belonging to trigonometric and
rational class  are generated from (\re {nlslq2}) and its limiting form 
({\ref{LK}) respectively,
and therefore
 inherit       the same corresponding  $R$-matrices 
(\re{trm}) and (\re{rrm}) .

\subsection{Fundamental and regular models}
The Lax operator $L_{aj}(\la)$ in general
 acts on the product space $V_a \otimes h_j$, of  the common auxiliary
space $V_a$  and   the quantum space $h_j$ at site $j$. The models 
with   $V_a$  isomorphic to all
$h_j , j=1,\dots N$  and given by the fundamental representation
 are called {\it fundamental} models.
  For such models the finite-dimensional  matrix representations of  
   auxiliary and  quantum spaces  
 become equivalent  and  may lead to  
 $L_{al}(\la)\equiv R_{al}(\la).$ 
 However
for clarifying a misconception  
prevelant in the literature,
 we should stress that  a model is 
 represented  by its Lax operator only,  
the associated
 $R$-matrix accounts for the commutation property of the elements
of this  Lax operator
through QYBE. Therefore  even for a fundamental model the Lax
operator may differ from its $R$-matrix. 
 We may however demand for the fundamental models
 an useful additional 
property:
$L_{al}(0)=P_{al}$, known as the {\it regularity} condition 
 given through the permutation operator, which may be
 expressed  in the general case 
 as a $n^2 \times n^2$ matrix $P^{(n)}_{al} =\sum_{{\beta \al}=1}^n
 E_{\al \beta}^aE_{\beta \al}^l,$ 
$E_{\al \beta}$ being a matrix with its $({\al ,\beta}) $ element  
as $1$ and the rest $0$.

Recall that  the global operator $\tau(\la)$ is constructed
from the local Lax operators  $L_{aj}(\la),~j=1,\ldots N$ as  
$
\tau (\la)=tr_a\l(L_{a1}(\la)\ldots  L_{aN}(\la)\r),
$
where the transfer matrix 
$\tau (\la)$ acts on the total quantum space ${\cal H}=
\otimes_{j=1}^N
h_j. $ 
Therefore, from the knowledge of Lax operators it is possible to derive
all conserved quantities including the Hamiltonian, which  in
general would be nonlocal objects.  The above
  regularity condition on 
  Lax operators however allows to  overcome this difficulty
and obtain Hamiltonians with  nearest-neibour (NN) interactions.
Let us pay  some special attention to this 
 specific  group of models
since,
 as we will see below, the most  
important integrable models applicable to condensed matter physics
 related problems
are given by the regular models with $n=2,3,4$ etc.
For this reason,
 though   our main concern in this paper is
 $2\times 2$ auxiliary matrix space,  we describe here 
more general $n$ cases and demonstrate that Hamiltonians 
  of such different physical models   interestingly  have similar 
 forms,
 when expressed through 
the permutation operator $P_{jj+1}$. Such a permutation operator
  exhibits  space  interchanging
 property $P_{aj}L_{ak}
=L_{jk}P_{aj}$,  along with  \   $P^2=1$ and $tr_a(P_{aj})=1$.
For all regular and periodic 
 models,  using the freedom of cyclic rotation of matrices under the 
trace, we can express  the transfer matrix  as
\be
\tau (0)=tr_a\l(P_{aj} P_{aj+1} .. P_{aN}P_{a1} ..  P_{aj-1}\r) 
\ =(P_{jj+1} .. P_{jN}P_{j1} ..  P_{jj-1})tr_a(P_{aj}) 
\ll{tau0}\ee
for any $j$ and as its
 derivative with respect to $\la$  we similarly get
\be
\tau' (0)=tr_a \sum_{j=1}^N 
\l(P_{aj} L'_{aj+1}(0)  .. P_{aN}P_{a1} ..  P_{aj-1}\r) \ \ = \
\ \sum_{j=1}^N(L'_{jj+1}(0)  .. P_{jN}P_{j1} ..  P_{jj-1})tr_a(P_{aj}) ,
\ll{tau1}\ee
where  the periodic boundary condition: $
 L_{aN+j}= L_{aj} $ is assumed. Defining now   
 $H= c_1= {d \ov d \la} \ln \tau(\la)_{\mid \la=0}
 = \tau' (0)\tau^{-1}(0)$
   and  using (\re {tau0}),  
(\re {tau1}) we may construct the related Hamiltonian as 
\be
H= \sum_{j=1}^N \ L'_{jj+1}(0) \  P_{jj+1}
\ll{htau01}\ee
with only NN interactions, where all  nonlocal factors
are  canceled out due to relevant properties of the permutation operator.
Similarly taking higher derivatives of $\ln \tau(\la)$,  higher conserved
quantities $c_j, j=2,3,\ldots N$ can be   constructed for these regular
 models. 
Note that the conserved operator $c_j$
involves  interactions of $j+1$ neighbors. 

 For the simplest case of   $n=2$, we may take the Lax operator 
as the $R$-matrix given by 
(\re{R-mat}), which  satisfies clearly 
 the regularity condition $L(0)\equiv R(0)=P$ for both
trigonometric and  rational cases. Moreover for (\re{trm}) the part 
$L'_{jj+1}(0)$  in (\re{htau01}) introduces anisotropy
reproducing the  Hamiltonian of the  $XXZ$ spin chain (\re{XXZ}).
 However since for the rational case (\re{rrm})
     $L'_{jj+1}(0)=1,$  using
 the expression  $ P^{(2)}_{jj+1}=\sum_{{\beta \al}=1}^2
 E_{\al \beta}^jE_{\beta \al}^{j+1}\equiv \ha ( H^{\si}_{jj+1} +1), $ where 
$ H^{\si}_{jj+1}
=
\vec \sigma_j\vec \sigma_{j+1},$
  (\re{htau01}) is reduced clearly to 
the isotropic spin-$\ha$   Hamiltonian $H^{\si}=\sum_j H^{\si}_{jj+1}$ (\re{XXX}).
 
It is intriguing to note that for the rational class 
the same form of Hamiltonian 
$H= \sum_{j=1}^N P^{(n)}_{jj+1}$  with several   
 higher values of $n$
describes most of the important integrable models, though their physical
forms are given mainly through
various representations of the permutation operator.
For example, for $n=3$ corresponding to $SU(3)$ group
we can  express the permutation operator 
$P^{(3)}_{jj+1}=\sum_{{\beta \al}=1}^3
 E_{\al \beta}^jE_{\beta \al}^{j+1} $ through spin-1 operators
$\vec S$ giving a variant of the integrable spin-1 model
\be
H= \sum _j \vec S_j\vec S_{j+1} + \epsilon
( \vec S_j\vec S_{j+1})^2, \ \ \mbox {with}  \ \epsilon=+1. \ll{s1}\ee
Considering a supersymmetric invariant
  $gl(1,2)$ case, i.e. realizing 
the corresponding graded permutation operator 
 $ P^{(1,2)}_{jj+1}$ using fermionic $( c_{aj},
c^\dagger_{aj} ), a=\uparrow,\downarrow$ and spin $\vec S$ operators
we may construct again from the Hamiltonian density
$ H^{tJ}_{jj+1}= (2 P^{(1,2)}_{jj+1}-1) $ the well known integrable $t-J$ model \c{tj}
\be \ H^{tJ}=\sum_j H^{tJ}_{jj+1}= \sum_j 
  -t{\cal P}\left(\sum_{\sigma=\uparrow\downarrow}
	c^{\dagger}_{\sigma j} c_{\sigma j+1} +h.c.
   \right){\cal P}+\
  J\left({\bf S}_j{\bf S}_{j+1} 
  -{1 \over 4} {n}_j{n}_{j+1}\right)
  + {n}_j+{n}_{j+1} \ \ll{tJ}\ee
with $J=2t=2$, where ${\cal P} $ projects out the double occupancy states.

A different $4$-dimensional realization of the fermion 
operators on the other 
hand  converts the same Hamiltonian to an integrable  correlated electron
model proposed in \c{links}.

Similarly 
for $n=4,$ i.e. for $SU(4)$,
realizing 
 $
P^{(4)}_{jj+1}
=P^{(2)}_{jj+1}\otimes P^{(2)}_{jj+1}$ in the factorized form  we can  get 
\be \  H=\sum_j P^{(\si)}_{jj+1}\otimes P^{(\tau)}_{jj+1}= 
{1 \ov 4} \sum_j(
H^{\si}_{jj+1}+1)( H^{\tau}_{jj+1}  +1 )=
{1 \ov 4} [H^\si+H^\tau+\sum_j
(H^{\si}_{jj+1} H^{\tau}_{jj+1}  +1) ] \ , \ll{sladder} \ee
with $ H^{\si,\tau}$ representing isotropic spin-$\ha$ Hamiltonians (\re{XXX}).
Adding now interaction along the rung:  $
H_{rung}=J\sum_j \vec \sigma_j \vec \tau_j , $ 
with $\ \ [H,H_{rung}]=0 $ to (\re{sladder}), where   $\sigma, \tau$
represent the spins along two legs of the ladder,
  we may  construct a model which is 
nothing but the integrable spin-$\ha$ ladder
 discovered recently \c{wang}. 

On the other hand,  from the same form of Hamiltonian but by
considering a supersymmetric extension $SU(2,2)$
 we may  realize
 $
P^{(2,2)}_{jj+1}
$ again through  fermion operators $( c_{aj},
c^\dagger_{aj} ), a=\uparrow,\downarrow$,  to
 construct an integrable extension of the  Hubbard
model proposed in \c{koress}.

 One can  repeat  the above construction of the 
spin-ladder model  for generating also  an integrable $t-J$ ladder model
introduced in \c{kunfrahm}, which would therefore 
corresponds   to a similar
 construction in  $n=6$ with Hamiltonian \bea
 H&=& \sum_j P^{(2,4)}_{jj+1}=\sum_j P^{(1,2)}_{jj+1} P^{(1,2)}_{jj+1}=
{1 \ov 4} \sum_j(
H^{(1)tJ}_{jj+1}+1)( H^{(2)tJ}_{jj+1}  +1 ) \nonumber \\ &=&
{1 \ov 4} [H^{(1)tJ}+H^{(2)tJ}+\sum_j
(H^{(1)tJ}_{jj+1} H^{(2)tJ}_{jj+1}  +1 )], \ll{tjlad}\eea 
where $H^{(a)tJ}, a=1,2$ are $t-J$ Hamiltonians (\re{tJ})
 along two legs. Adding 
a suitable $\ H_{rung} $ to (\re{tjlad})
 with $ \ [H,H_{rung}]=0, $ defining the interaction along the rung
  we finally obtain the integrable $t-J$ ladder model. 
 
Apart from the above applications of integrable systems  having similar
structure from the algebraic  point of view, we should mention  some
other important models like
 Hubbard model and the Kondo problem, which  also falls in the class of 
exactly solvable problems in one-dimension  \c{skryabin}.
Employing further twisting and gauge transformations on  multi-fermion
or multi-spin  integrable models one can generate another 
type of integrable models 
of current interest \c{kuntwist}. 
 Importance
 of solvable models in  physical systems,
 their relevance to experiments and  related isuues 
are discussed in  \c{indrani-angel}. For  detailed
and involved application of
Bethe ansatz technique 
 including that for  the theory of correlation
functions to various  integrable systems like $\delta$-bose gas,
 NLS, sine-Gordon etc. the readers are
referred to \c{korbook}.
\subsection{Fusion method}
We have constructed  spin, boson as well as the q-spin and q-boson  
  models through realization
of particular  Lax operators with inequivalent auxiliary and quantum spaces.
However in case of  finite-dimensional higher rank  spin representations 
there exists an intriguing method, known as the {\it fusion method},
 for obtaining higher  spin
models by { fusing} the elementary $R$-matrices like (\re{R-mat}).
 Thus by  fusion of only the quantum spaces one  can construct 
spin-s Lax operators with a spin-$\ha$ auxiliary space, as obtainable 
also directly from  (\re{nlslq2}) as a particular realization.
   Fusing further the
auxiliary spaces 
 the higher-spin  Lax operator with spin-s  auxiliary space may be
constructed as
\be
L_{{\bf ab}}=(P^+_{\bf a }\otimes P^+_{\bf b })
 \prod_{j=1}^s \prod_{k=1}^s R_{a_jb_k}(\la+i\al
(2s-k-j))(P^+_{\bf a } \otimes P^+_{\bf b }) g_{2s}(\la) \ll{fuse} \ee
with  $P^+_{{\bf a}({\bf b}) } $ as
 the symmetrizer in the fused spin-s space 
${\bf a } ({\bf b} ) $ and $g_{2s}$  some normalizing factor \c{babus}.
For the rational $R$-matrix corresponding to  (\re{rrm}) one obtains from 
(\re{fuse})
  the 
 integrable spin-s
 Babujian-Takhtajan  model \c{babus}, which for $s=1$ may be
 given in the same form as  Hamiltonian (\re{s1}), but with $\epsilon =-1$.
Similarly for the trigonometric case (\re{trm}) the fused model would
correspond to integrable anisotropic higher-spin chain.

It may however be stressed that  such 
fusion technique, as far as we know,
 has not been formulated yet for  
   bosonic and q-bosonic models. Such extension , at least for
the restricted values of $q$, needs therefore more attention.
\subsection{ Construction of classical models}
The  systematic  procedure   for constructing  quantum integrable models 
as various reductions of the same ancestor model,  
as described here, is applicable naturally also 
  to the corresponding classical models by taking the
 classical limit $\hbar \rw 0$. At this limit all 
 field operators would be transformed
to ordinary functions with their commutators reducing to the Poisson
brackets.  Note also that parameter $\al$ appearing in  
 the $R$-matrix   is   scaled actually as 
 $\hbar \al$, which  
yields the  classical $r$-matrix: $R(\la)=I +\hbar r(\la)
+O(\hbar^2)$ and reduces  QYBE (\re{ybe}) 
 to its classical limit $
 \{L_{ai}(\la),  
L_{bj}(\mu)\}= \delta_{ij}[r_{ab}(\la-\mu),   L_{ai}(\la)L_{bj}(\mu)] 
$.
The  classical   Lax operator reduced from (\re{nlslq2}) 
 would remain
however  almost in  the same form, though 
 the corresponding quantum algebras would change    
 into their corresponding  Poisson algebras.
This aspect of classical integrable systems is given in great detail in the
excellent monograph \c{Hamil-ft}.
 Using  these  classical analogs of quantum systems 
 one can   therefore apply the   algebraic 
 scheme  formulated above for  generating
  quantum  models also in classical context
and construct systematically 
the corresponding classical integrable models \c{classm}.

\section{Integrable statistical systems: vertex models}
\setcounter{equation}{0}
$D$-dimensional quantum systems are  known to be related to $(1+D)$-dimensional
classical statistical models, which is true   naturally also
 for $D=1$, where the  integrability of models might get manifested.
Interestingly, integrable quantum spin chain
and the corresponding  vertex model
 share  the same quantum $R$-matrix and  have  the same  
 representation for the transfer matrix,   
  commutativity of which:  $[\tau(\la),\tau(\mu)]=0 $  
 guarantees their integrability.
However, while the  spin chain
Hamiltonian $H_s$  is expressed through the transfer
 matrix  as $ ln \tau(\la)= I + \la H_s + O(\la^2)$,  the partition
function $Z$
 of the  vertex model is constructed from  $\tau(\la)$ as $Z=
tr(\tau(\la)^M)$. 
The known integrable  vertex models are usually related to the
quantum fundamental  models described above.

In conventional   vertex models 
 each bond connecting $N \times M$ arrays in a 
2-dimensional lattice  can take $n$ different possible random  values 
with certain probabilities, which
 for a
 configuration $i,j;k,l $ of bonds  meeting at each vertex point
 is  given by the Boltzmann weights
$w_{ij,kl}$.
These Boltzmann weights may be assigned as matrix elements 
$w_{ij,kl}(\la)=R^{ij}_{kl}(\la)$
 of a $R$-matrix (though 
 it might be of a more general $L$-operator,
as we will see below), which for integrable models must
satisfy the Yang-Baxter equation (\re{ybe}) and correspond   to 
 a  quantum integrable model.
The  partition function  of these vertex models may be    expressed  as 
 $Z= \sum_{config} \prod_{a,b,j,k} \omega_{a,j;b,k}(\la).$

The  simplest    among the vertex models  for  $n=2$
is the {\it 6-vertex model}
 \c{baxter},  which corresponds to the $XXZ$ spin chain 
 and may be 
 defined on a square lattice with a random direction on
 each bond ( left or  right on the horizontal, up or down on the vertical), 
constrained by the {\it ice rule},
 that the number of incoming and outgoing arrows at each vertex are the
same. This leaves only $6$ possible configurations and the corresponding 
 Boltzmann weights may be given by $6$ nontrivial matrix elements 
of the $R$-matrix (\re{R-mat}) with (\re{trm}). 
 It is fascinating that this model may
describe the possible 
configurations of Hydrogen (H) ions around Oxygen (O) atoms in an ice
crystal having
 two  different ({\it close-removed}) positions of the
H-ions relative to the O-atom in the H-bonding, while the ice rule
corresponds to the charge neutrality of the water molecule.

 A more { general  6-vertex model} may be obtained if instead we assign
its Boltzmann weights  directly to the spin-$\ha$ matrix
representation of the general ancestor 
Lax operator (\re{nlslq2}).  The 
 parameters $c_1^+=-c_1^-=\rho_+, \ c_2^+=-c_2^-=\rho_-$
 present in the Lax operator may be combined to serve  as the  
 horizontal $h \sim \ln \rho_+\rho_- $ and vertical $v \sim 
\ln {\rho_+ \ov \rho_-} $ fields acting on the model, which
   recovers amazingly the
most {\it general 6-vertex model}  proposed many years ago 
  \c{yys} through a  different construction. This also confirms the fact
that the Lax operator 
 (\re{nlslq2}) is indeed in the core of integrable quantum as
well as statistical models. 
 Using twisting  transformation  one can recover also  the  
 $6V(1)$ vertex model introduced in \c{wadati}.

We may consider higher vertex models with $n >2$, which may be obtained 
from the $R$-matrix (or the Lax operators) of the corresponding 
quantum integrable fundamental models with higher-dimensional auxiliary
spaces.
The well known examples are
the $19$-vertex model \c{19v} related to the
 Babujian-Takhtajan integrable spin-1 model 
\c{babus}, the Boltzmann weights of which 
may be given by the matrix elements of  Lax operator
 (\re{fuse}) with $s=1$. Similarly one may construct   the 
vertex models
  equivalent  to the Hubbard model,
  supersymmetric t-J model, 
 Bariev chain etc. 
 \c{vertex}.

In a following section
  new type of vertex models will be constructed 
from  our ancestor Lax operator using
 nonfundamental representations.

\section{Directions for constructing new classes of ultralocal models}
\setcounter{equation}{0}
 The same unified  scheme described in sect. 3 for 
constructing  integrable models
may be used also to indicate  various directions for  generating new 
integrable classes  
of quantum and statistical  models.

\subsection{Inhomogeneous  models}
In all 
above constructions  the central elements
in  the ancestor models 
(\ref{nlslq2}) or   
(\re{LK}) are chosen  as constant parameters. However 
if they are chosen  as  site dependent (or may even time dependent)
functions we can get an {\it inhomogeneous} class of models.
In these cases  the $c$'s 
would be  attached with site indices  as $ c_{j}'$s in the Lax operators and similarly 
 in  (\re{g}) $ M^\pm_j$ would appear as 
 functions, leading to the corresponding 
 inhomogeneous extensions of the known integrable
  models, namely  inhomogeneous lattice sine-Gordon,
Liouville, Toda chain, NLS model etc.
 However
since the local algebra remains  same as the original model, 
they have the same quantum
$R$-matrices.
Though similar inhomogeneous Toda chain, NLS models etc.  were proposed
earlier as classical systems, they seem to be
 new and yet unstudied as quantum models.  Recall 
that  the  impurity models proposed earlier \c{impure}
 fall into this class and 
are obtained by a particular choice of inhomogeneous $c_j$'s
which amounts to a shifting of the  
spectral parameter. Implementing the same idea to 
  the $XXX$ spin chain we notice that, if  in its constructing  
along with $c^0_a=1$ we choose  $c^1_2=-c^1_1=\epsilon_j$ resulting again 
 $m^+=1, m^-=0$, we get the same  
 form of the Lax operator, but with a shift $ L_j(\la-\epsilon_j)$,
resulting  that of the Gaudin model (\re{gaud}). Similarly higher spin
representations
as well as  $su(1,1) $ 
variant  would yield  other generalizations  of the 
  same model.  The commuting set of
Hamiltonians for the Gaudin model  may be  
 generated from its transfer matrix  at the limit $\al \to 0$ \c{gaudin} as  
$H_j={1\ov \al ^2 (\prod_k^N (\la-\epsilon_k))}\tau(\la \to \epsilon_j)
, \ j=1,2, \ldots, N$. Remarkably, the  Gaudin model may be mapped into
the  integrable BSC model, which is of immense contemporary interest
 \c{angela}.
  
 Physically such inhomogeneities may be interpreted as 
impurities, varying  external fields,  incommensuration etc.

\subsection{Hybrid models}

Another way of 
constructing  new  models  
 is to  use different realizations of
 algebras (\ref{nlslq2a}) or (\ref{k-alg}) at different
lattice sites,
depending on the type of the $R$-matrix.
For example  one may  consider spin-$\ha$ and spin-1
representations of  $su(2)$ at alternate lattice sites, which 
was  realized  actually in \c{alternates}.
 However we can  build   more general inhomogeneous
 integrable models by considering 
  different
underlying algebras and  different Lax operators at differing sites.
The basic idea is that the Lax operators representing  different models that
 are descended from the same
ancestor model 
and  share the same $R$-matrix can be combined together to build 
various hybrid models preserving quantum integrability.
For example,
we may consider  fermion-boson or spin-boson interacting models by inserting
alternatively spin-$\ha$ and bosonic (or q-bosonic) Lax operators at
alternate sites. One of such  physical constructions would be the celebrated
Jaynes-Cummings model.  
 It is  possible also  to construct some exotic hybrid integrable 
models, an example of which could be
 a  hybrid  sine-Gordon-Liouville model, where for $x \geq 0
$ it would follow
the sine-Gordon dynamics, while for $x <0$ the Liouville dynamics!


\subsection{Nonfundamental statistical models}
Vertex models, as mentioned, are   described  generally 
by the  $R$-matrix of a
regular quantum integrable model.
 However 
one can  construct 
a new class of integrable vertex models by exploiting 
a richer variety of nonfundamental 
 systems, where we  
 define the Boltzmann weights as  matrix  elements of the 
generalized   Lax operator (\re{nlslq2}): 
 $L_{ab}^{j,k}(u) =\omega_{a,j;b,k}(u)$,  with the 
use of  the explicit matrix representation for the basic operators
$S^\pm,S^3$ as
  \be
<s,\bar m|S^3|m,s>=m \delta_{m,\bar m},  \quad <s,\bar m|S^\pm|m,s>
= f^\pm_s(m)\delta_{m\pm 1,\bar m} .\ll{qsrep}\ee 
 Here $f^+_s(m)=f^-_s(m+1) \equiv g(m)$ is defined as in (\re{g}).
 Such general
 Boltzmann weights would now represent an ancestor  vertex model
  analogous to the quantum case and would  generate 
 through various reductions   new series of  vertex  models, 
 linked to
 q-spin and q-boson with generic q, q roots of unity and $q
\to 1$ \c{kuna02}.
 In all these models, generalizing the usual approach
 the horizontal (h)  and vertical (v) links 
may  become  inequivalent and independent  at every vertex point.  
 The h links, which are related to the auxiliary space    admit
 2 values, while 
 the v links, which correspond to the  quantum space  may  have
richer  possibilities with $j,k \in [1,D] $,
  $D$ being the dimension of 
the nonfundamental  matrix-representation of the  q-algebras.
The familiar ice-rule is generalized here as the 'color' conservation 
 $a+j=b+k$ for determining nonzero  Boltzmann weights.
Note that, alternatively finite-dimensional higher 
spin and $q$-spin vertex models
 can also be constructed  using the
 fusion  technique \c{babus}.

An interesting   possibility of regulating dimension  for the matrix
representation opens up at
$q^p=\pm 1$, when a
  variety of new q-spin and q-boson vertex models with finite-dimensional
representation  can be generated \c{kuna02}.

As in quantum models we can also construct here  
 a rich collection of { hybrid models} by combining different
 vertex models of the same class and inserting 
their defining  Boltzmann weights
along the vertex points $l=1,2,\ldots,N$ in each row,
 in any but in the same  manner. 
Due to the  association 
with the same $R$-matrix the integrability of such statistical
models is naturally  preserved.

\section { Unified   Bethe ansatz solution}
\setcounter{equation}{0}
In physical models our  aim  usually is to solve the eigenvalue problem  
for the Hamiltonian only.  
Solvable models allow such   exact solutions  $  H\mid m>= 
  E_m\mid m> $ through
coordinate formulation of the Bethe ansatz (CBA)  \c{bethe31}, which
 was used successfully in many  condensed matter
physics related problems like spin-chain, attractive and repulsive 
$\de$-Bose gas, Hubbard model etc \c{condmat}.
 Nevertheless CBA depends heavily on the structure 
of the Hamiltonian  of individual models and lacks consequently 
  the unified approach
of its algebraic formulation. 
We would  focus here 
 briefly only  on the   algebraic Bethe ansatz (ABA)  
 \c{aba,korbook}, which  under certain conditions can solve
 the  eigenvalue problem 
for the spectral parameter-dependent transfer matrix 
$  \tau(\la)\mid m>= 
  \Lambda_m(\la)\mid m> $
 and hence through its expansion  the eigenvalue problem for the whole
  set of conserved  operators, simultaneously.
Moreover, the ABA due to its predominantly
 model-independent features, which we will demonstrate below, 
appears to be a fairly   universal method.

Since the eigenvectors are common for all commuting conserved operators,
 by   expanding 
$\ln \Lambda_m(\la)$ simply as
\be
  c_1 \mid m>= 
  \Lambda_m'(0)\Lambda_m^{-1}(0)\mid m>
,~~c_2 \mid m>= 
  (\Lambda_m'(0)\Lambda_m^{-1}(0))'\mid m>
\ll{evpcn}\ee
etc. we obtain their   respective values,
 where   one may take $H=c_1$ or other combinations of 
$c$'s as the Hamiltonian, depending on the concrete  model.
This  powerful   method applicable to both  integrable   
   quantum and statistical  systems
requires however
 explicit 
knowledge of the associated Lax operator and 
 the $R$-matrix.

It  may be noticed that the {\it off}-diagonal element $B(\la)$ ($C(\la)$)
of the monodromy matrix (\re{monod}) acts generally 
 like  creation (annihilation) operator for the pseudoparticles,
 induced by the local
creation (annihilation) operator as the matrix elements in 
 $L_j(\la)$ acting on the quantum space at $j$. Therefore
the $m$-particle state $\mid m>$ may be  created 
by  acting $m$ times with $B(\la_a)$ on the pseudovacuum $\mid 0>=\prod_j^N\mid
0>_j$ , giving
$ \mid m>= B(\la_1) B(\la_2)\ \cdots B(\la_m)\mid 0>
$, where we suppose the crucial annihilation condition $C(\la_a)\mid 0> =0$.

Now for solving the eigenvalue problem of  $
\tau(\la)=A(\la)+ D(\la)$ exactly, we have to drag this operator
 through the string of $B (\la_a )$'s without spoiling their structure 
and  finally hit the 
pseudovacuum giving $A(\la)\mid 0>=\al(\la)\mid 0>
$ and $
D(\la)\mid 0>=
\bet(\la)\mid 0>$.
 For this purpose therefore one  requires 
  commutation  relations between the
elements  of (\re{monod}), which  for ultralocal models  may be derived 
from the QYBE (\re{gybe}). This, apart from  ensuring the  integrability
 of the system, is another 
important role played by  (\re{gybe}),
 yielding 
the relations
\bea
A(\la) B(\la_a)&=&  f(\la_a-\la)  B(\la_a) A(\la)  - f_1(\la_a-\la)
  B(\la) A(\la_a), \nonumber \\
D(\la) B(\la_a) &=& 
 f(\la-\la_a) B(\la_a) D(\la) -  f_1(\la-\la_a)  B(\la) D(\la_a), \ll{db} 
\eea
together with the trivial commutators   $ [A(\la), A(\mu)]  = 
[B(\la), B(\mu)]  = [D(\la), D(\mu)]  = [A(\la), D(\mu)]  = 0$ etc.,
where $f(\la)={a(\la) \ov b(\la)}$, $f_1(\la)={c(\la) \ov b(\la)}$ 
are combinations of the elements
from the    $R(\la)$-matrix  (\re{R-mat}).
We notice  that
(\re {db})   
are almost the  right kind of relations but for the second
terms in both the RHS, where the argument of $B$
  has changed
spoiling the structure of the eigenvector. However, if we put  the
sum of all such {\it unwanted} terms $= 0,$ we should
 be able to achieve our goal. 
In field models such unwanted terms   vanish automatically,
 while in lattice
models
their  removal amounts to
 the  Bethe equations, which may be induced independently  by
the periodic boundary condition, giving
\be 
\l({ \al(\la_a) \ov \bet(\la_a)}\r)^N=
 \pr_{b \not =a}{ f(\la_a-\la_b) \ov
 f(\la_b-\la_a)}
, ~~ a=1,2, \ldots ,m.
\ll{be}\ee
 Therefore the ABA  finally solves the eigenvalue problem for $\tau(\la)$
 yielding
\be \Lambda_m (\la) =
 \left(\pr_{a=1}^m f(\la_a-\la)\right)
\al(\la)+
\left(\pr_{a=1}^m f(\la-\la_a)\right)
\bet(\la),
\ll{lambda}\ee
where the Bethe equation (\re{be}),
 which is equivalent also to the singularity-free condition
of the eigenvalue (\re{lambda}) serves  in turn  as the set of 
   equations for determining the parameters $\la_a$.

Note that in both the above equations 
$\al(\la)=(<0|\hat L^{11}_j(\la)|0>)^N$ and 
$\beta(\la)=(<0|\hat L^{22}_j(\la)|0>)^N$ are the only 
 model dependent parts given by 
   the action of the upper and lower
diagonal operator elements $\hat L^{ii}_j(\la), i=1,2$ of the  
Lax operator of the model on the
pseudovacuum.
For  vertex models, for which  the ABA formulation goes parallelly,
  the Lax operator elements 
 in the above equations  
should be replaced by their matrix representations expressed through 
the  Boltzmann weights  as
$ 
<0|\hat L^{11}_j(\la)|0>= \omega_{+,1;+,1} (\la), \ \
<0|\hat L^{22}_j(\la)|0>=\omega_{-,1;-,1} (\la)
$.
It is  remarkable that 
  the rest of the  terms in  (\re{lambda}) and (\re{be}) 
are  given solely through the $R$-matrix elements 
 $f(\la)$ and therefore depend only on
 the related class  (\re{trm}) or (\re{rrm}).
Recall that in  integrable  models, as described in sect. 3, 
the $R$-matrix remains same for all models belonging to a particular class,
while   the $L$-operators differ and  may be obtained 
 through various reductions from the
same ancestor Lax operator.

 Therefore  taking
the Lax operator elements in (\re{lambda}) and (\re{be}) as  
  those from the general Lax operator (\ref{nlslq2}), one may   consider  
the above  eigenvalue   and the Bethe equation  to be
 the unifying equations for      exact solution of 
all   integrable ultralocal
quantum and  statistical models 
constructed here.
Consequently models like the DNLS, SG, Liouville
and the $XXZ$ chain together with the 6-vertex model,
 belonging to the trigonometric
 class (\re{trm})  should share  similar  
eigenvalue relations with individual differences appearing only in 
the   form of  $\al(\la)$ and $ \bet (\la)$ coefficients.
Thus  
this  deep rooted
 universality  in integrable systems  helps to solve the eigenvalue
problem for the whole class of models and for the full  hierarchy of
their 
conserved currents in a systematic way. 
Let us present the explicit example of   $XXZ$ chain with 
 Lax operator  (\re{XXZ}),
 defining $\mid 0>$ as all spin up
state which gives
$\al(\la)=\sin ^N (\la+ \al), \ \bet(\la)=\sin ^N \la $
 in   Bethe equation (\re{be}) (with  a shift $\la \rw \la+{\al \ov 2} $)
resulting
\be 
\left(  {\sin (\la_a+{\al \ov 2}) \ov \sin (\la_a -{\al \ov 2})} \r)^N=
 \pr_{b \neq a}^m {\sin (\la_a-\la_b+\al) \ov \sin (\la_a-\la_b - \al)}.
\ll{bexxz}\ee
for $a=1,2, \ldots , m.
$   Similarly (\re{lambda}) gives  the eigenvalue 
\be \Lambda_m^{XXZ} (\la) =\sin ^N (\la+ \al)
 \pr_{a=1}^m {\sin (\la_a-\la+{\al \ov 2}) \ov \sin (\la_a-\la -{\al \ov 2})}
 +\sin ^N \la
\pr_{a=1}^m {\sin (\la -\la_a+3{\al \ov 2}) \ov \sin (\la-\la_a+{\al \ov 2})}
\ll{lamxxz}\ee
 yielding for 
 $H_{xxz}=c_1,$   the energy
spectrum
\be
  E_{xxz}^{(m)}= \Lambda_m (\la)'\Lambda_m^{-1} (\la)\mid_{\la=0}=
\sin \al\sum_{a=1}^m {1 \ov \sin (\la_a -{\al \ov 2}) 
\sin (\la_a + {\al \ov 2})}+ N\cot \al.  
\ll{exxz}\ee
At the  limit $\al \rw 0, \ \sin \la \rw \la$, when the $R$-matrix along
with its associated models reduce to the rational class, one can  
derive   
the corresponding Bethe ansatz results by taking the rational limit of the
above equations. For example the relevant equations  for
 the isotropic $XXX$ chain can be obtained directly from 
those  for the  $XXZ$ chain presented above.
 Intriguingly the corresponding result 
  for the  NLS lattice
model, which  belongs  to  the same  rational class,   
 should  also show close  similarity with that of the $XXX$ chain.


\section{Quantum 
integrable nonultralocal models}
\setcounter{equation}{0}
Though many celebrated classical integrable models like KdV, mKdV, nonlinear
$\sigma$-model, derivative NLS etc. belong to the class of nonultralocal
models, successful quantum generalization could be made only for  handful of
them.  The reason, as mentioned already, is the violation of the
ultralocality condition. Recall that 
this condition helps to transit from  local  QYBE to its
global form and consequently establish the integrability 
for  ultralocal systems. Therefore  
the key equations and the related 
  formulation for the integrability theory of 
the nonultralocal  
models  must be suitably modified.   
\subsection {Braided  extensions of  QYBE}
For understanding first the algebraic structures underlying the
 nonultralocal  systems 
we have to note that the  trivial multiplication property (\re{mult})
valid for ultralocal models   needs
to be generalized here 
as
  $
({A}\otimes  {B})(C \otimes D) =\psi_{BC} (A(C \otimes B)D)
$
where the braiding $\psi_{BC} $  takes into   account the
 noncommutativity of $B_2, C_1$. In spite of such braided extension  of the 
multiplication rule, the associated 
 coproduct structure  of   the underlying Hopf
 algebra,    crucial for transition to the  global QYBE,
 must be preserved. Such a 
 braided extension  of the Hopf algebra \c{majid,hlavqb} 
was implemented in formulating the integrability theory of 
nonultralocal models through an unified approach 
 \c{hlavkun96}.
The basic idea is to 
complement the commutation rule for the Lax operators at the same site
  with
their braiding property  at different
lattice sites. Note however that
 in general the braiding may  differ
widely  and with   arbitrarily  varying  ranges
 the  picture 
might become too complicated for explicit description.   
Therefore let us   limit
 first   to the nearest-neighbor (NN)  type braiding 
\begin{equation}
L_{2 j+1}(\mu) Z_{21}^{-1}L_{1 j}(\la) = L_{1 j}(\la)
 L_{2 j+1}(\mu)
 \ll{zlzl1u}\end{equation}
assuming that the ultralocality  holds starting from
the next neighbors. A pictorial description of this condition is given
in Fig. 1a).

The local QYBE at the same time  must also  be generalized to incorporate the  
 braiding relations, such that
 the transition  to its global form 
becomes possible again. 
Such  braided extension of the QYBE (BQYBE) compatible with (\re{zlzl1u})
 takes the form (see Fig. 1b) ) 
\begin{equation}
{R}_{12}(\la-\mu)Z_{21}^{-1}L_{1j}(\la)L_{2j}(\mu)
= Z_{12}^{-1}L_{2j}(\mu) L_{1j}(\la){R}_{12}(\la-\mu).
\ll{bqybel}\end{equation}
We list below the
known nonultralocal integrable models that
 can be described by the above braided
equations.
Note that  the quantum $ R$-matrix appearing here  is the same 
(\re{R-mat}) as for the ultralocal systems. However 
 the additional 
braiding matrix $Z$,  unlike the $R$-matrix  seems to be 
model-dependent and generally independent of the spectral parameter,
 though similar to the $R$-matrix it  satisfies the 
YBE like equations and   might also  
 become  spectral parameter dependent for specific models \c{hlavkun96}.

The next step is the global extension of the BQYBE for the monodromy matrix
(\re{monod}) and it is not difficult to check  that due to the braiding
relation (\re{zlzl1u}), the form of BQYBE is preserved for global matrices
like $T_{a}^{[k,j]}(\la)= \prod_{j=1}^kL_{aj}(\la)$ (see Fig. 1c)).
 However since for the periodic
boundary condition one imposes  $  L_{aN+1}(\la)=L_{a1}(\la)$,
the Lax operators $L_{aj}(\la)$
for $j=1 $ and $j=N$ again become  NN entries and hence modify the 
equation due to the appearance of an extra  
$Z$ matrix from the 
braiding relation (\re{zlzl1u}), leading finally to the global BQYBE
\begin{equation}
{R}_{12}(\la-\mu)Z_{21}^{-1}T_{1}(\la)Z_{12}^{-1}T_{2}(\mu)
= Z_{12}^{-1}T_{2}(\mu)Z_{21}^{-1} T_{1}(\la){R}_{12}(\la-\mu).
\ll{bqybeg} \ee

Though this equation is similar to (\re{gybe}),
the  commutation of the transfer matrices ensuring  the integrability 
of the systems through factorization of the trace identity
 becomes problematic due to the presence of $Z$-matrix.
Detail discussion of this problem and the classification of the $Z$-matrices 
allowing factorization    is given  in \c{hlavkun96}.
Investigations of some nonultralocal systems from  a different
 angle were done
in \c{reshet}.
It is easy to see that from 
the corresponding equations for the nonultralocal models
 presented above one can recover
 the   known relations for the ultralocal models by supposing 
 the braiding matrix  $Z=1$ (see also the caption in Fig 1)
\newpage
\epsfxsize=470pt \epsfbox{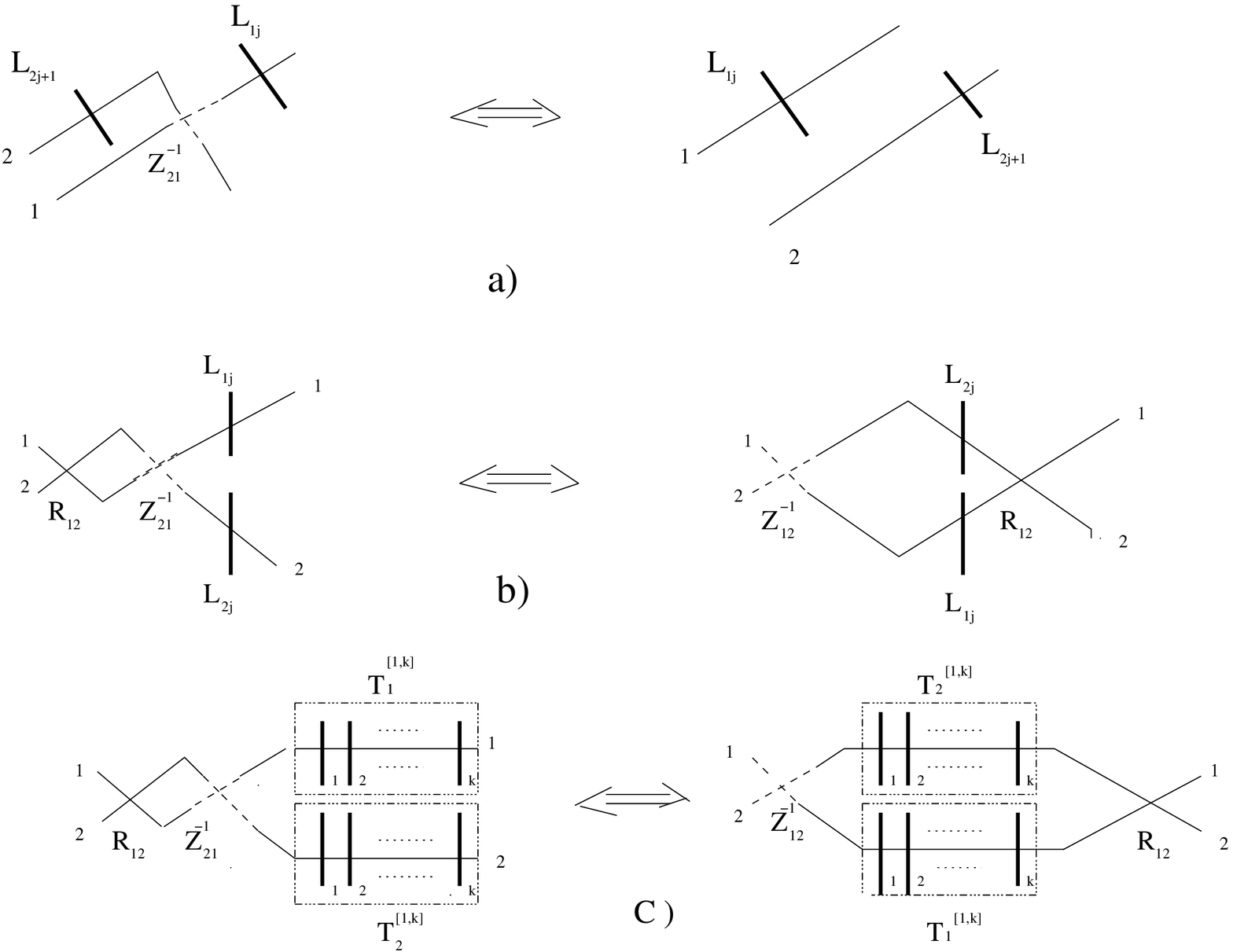}
\vskip 2cm
Fig. 1:
\ {\small Pictorial description of a) braiding relation (\re {zlzl1u}),
b) local braided QYBE (\re {bqybel}) for the Lax operators $L_{aj}(\lambda_a)$
and c) global     braided QYBE for
 $ T^{[1,k]}_a(\lambda_a)= \prod_{j=1}^kL_{aj}(\lambda_a), \  k<N$.  Note that 
putting $Z=1$, i.e. removing braiding by
 undoing the crossing of {\it dashed} lines in  above
figures 1a,b,c) 
one can recover the corresponding pictures  for the ultralocal models
\c{baxter}, namely
 ultralocality condition
(\re{ul}),
  local (\re{ybe}) and global  QYBE (\re{gybe}), respectively.} 
\newpage
\subsection {List of quantum integrable nonultralocal  models}
 Nonultralocal
 models are mostly nonfundamental systems with infinite dimensional 
representations defined  in some  Hilbert space.
They may correspond to 
  integrable models with spectral parameter dependent
 Lax operator and $R(\la)$-matrix or may describe only 
nonultralocal algebras   having
spectral parameterless $L$-operator  and  
$R(\la)_{\la \to + \infty} \to R^+_q $-matrix.  
Nevertheless the nonultralocal  quantum  models listed below 
should be   
described  through the same braided relations 
 (\re{zlzl1u},\re{bqybel},\re{bqybeg}) or their corresponding 
spectral-less form in a systematic way.
 Therefore we 
 present only   the  explicit form  
  of their  braiding matrix 
 $ Z$   and the $L$ operator, indicating  the class of  $R$-matrix they
belong to. These inputs should be enough to obtain 
all individual  equations and derive the related results.

\ni {\bf I. Systems with spectral parameterless $R$-matrix} 
\\ 
\ni  1. {\it Current algebra in WZWN model} \c{wzwn}
 
The model involves   the nonultralocal current algebra
\be \{L_1(x),L_2(y)\}= \frac {\gamma} {2} [C,L_1(x)-L_2(y)]\delta(x-y)
+ \gamma C \delta'(x-y) \ll{current}\ee
with $C_{12}= 2P_{12}-1$, where $P_{12} $ is the permutation operator, 
$L= \frac {1} {2} (J_0+J_1)$ with $J_\mu= \partial_\mu gg^{-1},$
is the  current and $g \in SU(N)$ the chiral field. 
Discretized and quantum version of this  algebra may be cast
 as the spectral-free limit of the above   braided YBE 
relations with
$R^+_q$ as the R-matrix, current $L$ as the Lax operator  
  and  $Z_{12}=R_{q21}^+$ as the braiding matrix, which takes the form
\begin{equation}
{R}^+_{q21}L_{1j}L_{2j}
= L_{2j} L_{1j}{R}^+_{q12}, \ \ L_{1j} L_{2j+1} = L_{2j+1}({R}^+_{q12})^{-1}
 L_{1j} 
\ll{wzwn}\end{equation}
For the details and an interesting quantum group relation of this model
the readers are refered to   the original works \c{wzwn}.
\\  2. {\it Coulomb gas picture of  CFT}  \c{babelon}

The Drinfeld-Sokolov linear problem : $ Q_x={\cal L}(x)Q$   
describing this system  may be given
 in the simplest case by the linear
operator $ {\cal L}(x)= v(x) \si ^3 - \si ^+ $ with a nonultralocal property
due to current-like relation ~$ \{ v(x) , v(y) \}= \delta'(x-y). $
Discretized and quantized forms
of the current-like operator defined through the commutation relations
\be
[v^\pm_k, v^\pm_l] = \pm i {\al \ov 2} (\delta_{k,l+1}- \delta_{k+1,l}),
\ \ 
[v^+_k, v^-_l] =i {\al \ov 2} (\delta_{k+1,l}-
2\delta_{k,l}+\delta_{k,l+1})
\ll{cr12}\ee
construct the corresponding discretized linear operator as 
$~~ L_k=e^{-i v^-_k\si ^3} + \Delta e^{i v^+_k} \si^+ ,
~~$, which  
similar to the above case satisfies  the  spectral-free    braided
YBE and other relations with $R^+_q$ as  $R$ and $Z=
q^{- \si ^3 
\otimes \si ^3}
$ as the braiding matrix. Generalization of this model 
for $SU(N)$ has also been 
constructed  similarly in   \c{babelon}.
 \\
\ni {\bf II. Models with rational  $R(\la)$-matrix}
\\
\ni 3. {\it Nonabelian Toda chain }  \c{natoda}

 The Lax operator of the model  given by
\be
L_{k}(\lambda) = \left( \begin{array}{c}
\lambda - A_k
\qquad  \ \ -B_{k-1}
 \\ I
\qquad \ 0
         \end{array}   \right), \ \ A_k=\dot{g}_k g_k^{-1},
 \ \  B_k= g_{k+1}
 g_k^{-1}, \ g_k \in SU(N), 
\ll{Lkor}\ee
represents nonultralocal integrable model and solves all braided relations 
including the BQYBE with spectral dependent 
 rational  
$~R(\lambda)= P-ih \lambda I$ and  the  braiding
matrix  $Z_{12}=
 {\bf 1} + i { h }(
 e_{22}\otimes e_{12}) \pi, $
where $P$ and $ \pi$ are permutation operators. For further details on this
model including its gauge relation with an ultralocal model we refer to the
original work   \c{natoda}.
\\  4. {\it Nonultralocal quantum mapping }\c{Nijhof}

The system is  
described by the  Lax operator
$L_n=V_{2n}V_{2n-1}$, with 
$ V_{n}=\lambda_n \sigma^- +\sigma^++ \ha v_{n} (1+\sigma^3),$
where  the  discretized operator  $ v_k \equiv v^-_k$
involves  nonultralocal algebra (\re{cr12}) and  
yields at the continuum limit  $\Delta \to 0$ 
 the current-like field: 
$v_k \to i \Delta  v(x)$. 
This nonultralocal quantum integrable 
model satisfies again 
  integrable braided  relations 
with  spectral-dependent
 rational $R(\la_1-\la_2)$-matrix  similar to  the above case but now
 with a spectral  dependent braiding matrix
 $\  \ Z_{12}(\la_2)= { I}- \frac { h }{\la_2}\sigma^{-
}\otimes \sigma^{+} \ \ $ and $Z_{21}(\la_1). $ For generalization of this
model to higher  rank groups and other details we refer again to the
original work  \c{Nijhof}.
\\ 
\ni {\bf III. Models with trigonometric $R(\la)$-matrix} 
\\
\ni 5. {\it Quantum mKdV model  }  \c{kmpl95}

This well known  nonultralocal model may
 be raised to the quantum level with discrete 
Lax operator
\be
L_{k}(\xi) = \left( \begin{array}{c}
(W_k^{-})^{-1}
\qquad  \ \
 i\De \xi W_k^{+}\\
-i\De \xi (W_k^{+})^{-1}
\qquad \ W_k^{-}
         \end{array}   \right),
\ll{Lmkdv}\ee
where $ W_{j}^{\pm}=e^{iv_{j}^{\pm} } $ with $v^\pm_k$ obeying the 
 nonultralocal 
 relations like (\re{cr12}).
  $R$-matrix (\re{trm}) and the braiding matrix
$ Z_{12}=  Z_{21}= q^{-\frac {1}{2} \sigma^3\otimes \sigma^3
}$ are associated with  this  nonultralocal integrable system 
\c{kmpl95}. Bethe ansatz 
solution  of quantum mKdV and its generalizations can be found in detail in
 \c{fr02}. It is seen easily that one  can recover the well known Lax
operator of the mKdV field model: $  U(x,\xi)={i \ov 2} (iv(x) \sigma^3
+\xi \sigma ^2)$  from (\re{Lmkdv}) at the field limit when $v^{\mp}_k \to 
\sqrt \mp  \Delta  v(x)$, as $L_k= {\rm I} + \Delta  U(x,\xi) + O (\Delta ^2)$. 
\\
\ni 6. {\it Quantum light-cone sine-Gordon  model } 

It is known that 
this well known equation : $ \partial^2_{+-}u=2\sin2 u$ may be represented 
by the 
zero curvature condition: \  $ 
\partial_- {\rm {U}}_+ -\partial_+ {\rm {U}}_+ + [ {\rm {U}}_+,
 {\rm {U}}_-]=0$ 
 of the 
 Lax pair ${\rm {U}}_\pm $ with 
$ {\rm {U}}_-(x) 
=
{i \ov 2} \partial_- u(x) \si ^3 +    
   \xi ( e^{-i  u (x)} \si ^+ +  
     e^{i  u (x)} \si ^-)
$ and similarly for ${\rm {U}}_+(x)$. Recently  quantum as well as  
exact lattice versions of the nonultralocal 
 Lax operator have been constructed \c{kun02}, which in particular for 
${\rm {U}}_-(x)$  may be given  in the form
\be
L^{(-)lcsg}_j{(\la)} =
   e^{i(p_j-\al \nabla u_j)\si ^3} + 
  \Delta \xi \left ( e^{-i(p_j+\al  u_{j+1})} \si ^+
    + e^{i(p_j+\al  u_{j+1})} \si ^- \right ) 
, \quad \nabla u_j \equiv  u_{j+1}- u_{j}. 
\ll{lcsg}\ee 
It may be shown also that (\re{lcsg}) 
  obeys exactly  the above BQYBE and the braiding
relation with the trigonometric 
$R$-matrix (\re{trm}) and the braiding
matrix 
$ Z_{12}^{(-)}=e^{i \al \sigma^3 \otimes \sigma^3} $,  and consequently
represent a genuine quantum integrable nonultralocal model.

Some other nonultralocal models known in the literature need
 introduction of braiding beyond 
NN, basic formulation of which can be found in \c{hlavqb,hlavkun96}.
 Examples of   
such models having same 
 braiding between any two different sites are
i.) {\it Integrable model on moduli space }  \c{alex},
ii.) {\it Supersymmetric models } \c{SUSY,hlavkun96}, 
iii. ) {\it Braided  algebra}  \c{majid}, iv) 
NUL extension of YBE \c{exybe}
 etc. Their unified description 
can be found  in \c{hlavkun96,review}. 

\subsection{Algebraic Bethe ansatz}

The solution of the eigenvalue problem for  integrable
nonultralocal models by diagonalizing
  the transfer matrix may  be
formulated through algebraic Bethe ansatz 
    exactly in analogy with the ultralocal models, whenever the
factorization of the trace problem, as mentioned above, could be resolved.
 The key equation that is to  be  used  
for nonultralocal models for finding the commutation
relations analogous to  (\re{db}) in the ABA scheme     
 should naturally be given by BQYBE (\re{bqybeg}). We however 
 skip all details of this
ABA formulation 
for nonultralocal models, which can be found in   explicit form on the example of 
 the
nonultralocal 
quantum mKdV model in  \c{kmpl95,fr02}.
\subsection{Open  directions in nonultralocal models}

Since some of the  nonultralocal models like nonabelian
Toda chain, WZWN current algebra, mKdV etc. described above can be connected 
to ultralocal models through operator dependent local gauge transformation,
it would be challenging to discover similar relation, if any,
for  the rest of the quantum integrable  nonultralocal models \c{kun02}.

Other challenging problems  undoubtedly are the possible 
 quantum integrable 
formulation of the famous nonultralocal models like nonlinear $\si$-model,
complex sine-Gordon model, derivative NLS equation etc.   through braided YBE.

As we know, there is a  remarkable interconnection between the integrable
quantum and statistical models. However this connection is discovered until
now only for the ultralocal models as we have also seen here.
 Therefore it should be 
a new direction of study  to investigate
 whether there could be any meaningful statistical model corresponding to
the integrable nonultralocal models described here.

Another problem worth looking into would be to formulate  fundamental 
nonultralocal models, if any,
 which then could be used possibly for generalizing   spin and
electron models with  nonultralocality.

Anyway since this vast branch of integrable systems has received
significantly
insufficient attention, we may hope to have many hidden excitements 
in this area.

\section {Concluding remarks}
\setcounter{equation}{0}
Quantum 
integrable systems can be divided into two broad classes, ultralocal (UL) and
nonultralocal (NUL). We have presented here a brief description 
 of such models 
with references for further details and 
  demonstrated that the models  belonging to both
these classes 
 can be described systematically through a set of algebraic
 relations signifying integrability of these systems. For UL models
 these relations are the ultralocality
condition and the QYBE involving Lax operator
 $L$ and the $R$ -matrix, while for
NUL models they are extended to braiding relation and braided QYBE with an 
additional entry of braiding matrix $Z$. The $L$ operator
representing an individual model is naturally model dependent and
the same seems to be
true also for the $Z$ matrix. The $R$-matrix on the other hand is mainly of
two types (elliptic case is not considered here),
 trigonometric and rational depending on the class of models
that are associated with q-deformed and undeformed algebras, respectively.
 This 
induces   a significantly model-independent  approach
also in the ABA method for solving the eigenvalue problem.
For UL systems, the theory of which is  more developed, 
 one can go beyond  and prescribe  an unifying algebraic scheme  for  
  generating   individual Lax operators 
 realized from  a single ancestor model in a systematic way.
It would be a challenge to extend the formulation of 
this scheme also for the NUL models.
The integrable statistical vertex models can be related to the corresponding 
quantum models, which as a rule belong to UL systems. Possible 
systematic extension of
such relation to NUL systems would be another challenging problem.

 \end{document}